\pdfoutput=1
\documentclass[reprint,aps,prd,amsmath,amssymb,superscriptaddress,floatfix]{revtex4-2}
\usepackage[T1]{fontenc}
\usepackage{lmodern}
\usepackage{booktabs}
\usepackage{multirow}
\usepackage{xcolor}
\usepackage{graphicx}
\usepackage{placeins}
\usepackage[colorlinks=true,linkcolor=blue!50!black,citecolor=blue!50!black,urlcolor=blue!50!black]{hyperref}
\newsavebox{\fitbox}
\newcommand{\fitwidth}[1]{\sbox{\fitbox}{#1}\ifdim\wd\fitbox>\linewidth\resizebox{\linewidth}{!}{\usebox{\fitbox}}\else\usebox{\fitbox}\fi}

\newcommand{\hook}[1]{\texttt{#1}}
\newcommand{\partkin}{\texttt{ParT\_kin}}
\newcommand{\partfull}{\texttt{ParT\_full}}
\newcommand{\pTD}{p_T D}
\newcommand{\mSD}{m_{\mathrm{SD}}}
\newcommand{\nconst}{n_{\mathrm{const}}}

\begin{document}

\title{What does the Particle Transformer know about QCD?\\
Sparse autoencoders and concept erasure with null controls}
\author{Neeraj}
\email{neeraj.neeraj@cern.ch}
\author{Amal Sarkar}
\email{amal.sarkar@cern.ch}
\affiliation{School of Physical Sciences, Indian Institute of Technology Mandi,
Kamand, Himachal Pradesh, India}
\date{November 26, 2026}

\begin{abstract}
Transformer jet taggers such as the Particle Transformer (\texttt{ParT})
outperform analytic jet-substructure observables, but which QCD information
they encode and use is not known. We put this question to the public kinematic
\texttt{ParT} checkpoint (\partkin) on JetClass with two machine-learning
interpretability tools, sparse autoencoders (SAEs) and linear concept erasure
(LEACE), and test every claim against null controls that can fail. As a
population, SAE latents of the trained model are far more often causal and
aligned with known jet observables than latents of an untrained network or of
shuffled activations, yet no individual
latent survives false-discovery-rate control on all three criteria: latents do
not replicate across SAE seeds, even with five times more training jets. At the
concept level, erasing all known observables
from the jet representation lowers the accuracy from $0.746$ to $0.413$, against
$0.699$ for random subspaces of the same rank, with claimed reliance on
infrared- and collinear- (IRC-) safe energy
flow polynomials, the Sudakov-safe ratios
$\tau_{21}$ and $\tau_{32}$, and the unsafe constituent multiplicity;
energy-matched ablation of groups of SAE latents supports reliance on jet mass
in every dictionary and SAE seed tested, and on $\tau_2$ in two dictionaries. Under exact collinear
splittings and soft emissions, however, the tagger's internal readouts of these IRC-safe and
Sudakov-safe observables move as much as its multiplicity readout: \partkin{} uses IRC-safe
information, but does not encode it in an IRC-safe way. The concept-erasure
and IRC results hold also for the Particle Transformer trained with particle
identification, and no SAE latent responds additively when a single particle is
split collinearly.
\end{abstract}

\maketitle

%==============================================================================
\section{Introduction}
\label{sec:intro}
%==============================================================================

At the Large Hadron Collider (LHC), heavy Standard Model resonances ($W$, $Z$,
and Higgs bosons, and top quarks) produced at high transverse momentum decay
into tightly collimated sprays of hadrons and leptons that are reconstructed as
single large-radius jets~\cite{Larkoski:2017jix}. Over the past fifteen years,
jet-substructure theory has developed a hierarchy of observables that separate
these multi-prong decays from the much larger single-prong QCD background: infrared- and collinear- (IRC-) safe quantities calculable in
perturbative QCD, such as jet invariant mass $m_J$, soft-drop groomed mass
$\mSD$~\cite{Larkoski:2014wba}, $N$-subjettiness
$\tau_N$~\cite{Thaler:2010tr} (whose ratios $\tau_{21}, \tau_{32}$ are not IRC
safe but Sudakov safe~\cite{Larkoski:2013paa,Larkoski:2015lea}), and the complete linear spanning
basis of Energy Flow Polynomials (EFPs)~\cite{Komiske:2017aww}, alongside
IRC-unsafe fragmentation variables such as constituent multiplicity $\nconst$
and transverse-momentum dispersion $\pTD$ that capture quark-versus-gluon colour
flow and hadronisation~\cite{Gallicchio:2011xq,CMS:2013kfa}.

Deep-learning jet taggers, in particular permutation-equivariant architectures
such as the Particle Transformer (\texttt{ParT})~\cite{Qu:2022mxj}, operate
directly on constituent four-vectors and outperform these analytic baselines by
wide margins; taggers of this kind are validated in collision data and used in
LHC analyses~\cite{CMS:2020poo}. Because their internal representations are opaque, it is hard to
tell how strongly a tagger relies on perturbatively controlled IRC-safe
substructure rather than IRC-unsafe hadronisation details, whether it exploits
simulation artefacts, or whether it computes new jet observables.
Earlier jet-interpretability work projected classifier outputs onto human-readable
bases~\cite{Faucett:2020vbu}, attributed scores to inputs and expert
variables~\cite{Agarwal:2020fpt}, or analysed attention maps and their
sparsity~\cite{Wang:2024interp,Legge:2025cnm}. More recently, mechanistic
interpretability methods have reached collider taggers via circuit analysis and
path patching of top taggers~\cite{Rai:2026uzm}, linear probing and ablations of
Lorentz-equivariant networks~\cite{Agarwal:2026uqw}, surrogate-based component
attribution on JetClass~\cite{Laatu:2026quw}, and resonance-steering features in
LHC generative models~\cite{Calderon:2026qqj}; ref.~\cite{Gambhir:2026interp}
reviews the conceptual landscape for physics.

In large language models, the standard tool for decomposing internal activations
into candidate concepts is dictionary learning with sparse autoencoders (SAEs),
which express activation vectors as sparse linear combinations of learned
directions~\cite{Cunningham:2023sae,Gao:2024sae}. In physics, SAEs have recently
been applied to an IceCube neutrino foundation model~\cite{BonnetGuerrini:2026sae}
and to neural quantum states~\cite{Qi:2026xmb}, and very recently to edit a jet
foundation model, where removing a few SAE latent directions repairs an
infrared-safety flaw~\cite{Bonnet-Guerrini:2026mwq}. To our knowledge
(INSPIRE and arXiv searches on 25 November 2026), no study has yet tested
whether the SAE features of a collider-jet tagger reflect learned physics
against null controls. Applying SAEs to a physics model
has a known pitfall: Heap et al.~\cite{Heap:2025random} showed that standard SAE quality and interpretability metrics do not distinguish
trained language transformers from randomly initialised ones. Any claim that an
SAE reveals \emph{learned physics} therefore requires null controls that can
falsify the method.

Here we treat the interpretability of a pretrained collider-jet transformer
(\partkin{} on JetClass) as a controlled measurement and address three physics
questions:
\begin{enumerate}
  \item \textbf{Dictionary atoms versus distributed subspaces:} Decomposed
    with top-$k$ SAEs under null controls and multiple-testing correction, does
    \partkin{} contain individually identifiable physics features, or only a
    population-level enrichment spread over redundant subspaces?
  \item \textbf{Causal reliance on IRC-safe and IRC-unsafe physics:} Using
    concept-level interventions that do not require identifiable single
    latents (linear concept erasure, LEACE~\cite{Belrose:2023leace}, and
    energy-matched ablation of groups of SAE latents), which IRC-safe and
    IRC-unsafe jet observables does \partkin{} causally depend on, and does it
    encode them in an IRC-safe way?
  \item \textbf{Completeness of known observables and hidden input cues:} How
    much classification power survives erasing all known observables, how is the
    remainder distributed over jet classes, and does the nominally
    ``kinematics-only'' tagger extract particle-identity cues from its inputs?
\end{enumerate}

To avoid post-hoc selection, we fixed the learned-feature criterion, the nulls,
the negative controls and the decision rules before looking at the
corresponding results. Appendix~\ref{app:protocol} lists every later amendment
and disclosure, all of which are recorded in a timestamped decision log.

Sections~\ref{sec:setup} and~\ref{sec:sae_framework} describe the setup and the
controlled SAE framework, Sections~\ref{sec:sae_results}--\ref{sec:irc} present
the SAE, concept-level and IRC results, Section~\ref{sec:full} repeats the
concept-erasure, IRC and mass tests on the Particle Transformer trained with
particle identification, and Sections~\ref{sec:discussion}
and~\ref{sec:conclusions} discuss and conclude.

%==============================================================================
\section{Collider physics setup, model geometry, and practical findings}
\label{sec:setup}
%==============================================================================

\subsection{JetClass topologies and reference observable basis}
\label{sec:data_obs}

We study the JetClass validation dataset~\cite{JetClass,Qu:2022mxj}, comprising
simulated proton--proton collisions (MadGraph5\_aMC@NLO~\cite{Alwall:2014hca} +
Pythia~8~\cite{Sjostrand:2014zea} + Delphes~3~\cite{deFavereau:2013fsa}
with a CMS-like detector card) clustered into large-radius anti-$k_T$ jets
($R=0.8$)~\cite{Cacciari:2008gp} with $p_T \in [500,1000]\,\mathrm{GeV}$ and
$|\eta| < 2$. The benchmark defines a balanced ten-class tagging problem
spanning distinct collider topologies (50 files of $10^5$ jets each):
\begin{itemize}
  \item \textbf{Single-prong background:} inclusive QCD quark/gluon jets;
  \item \textbf{Two-prong boosted bosons:} $W\to qq'$, $Z\to q\bar q$,
    $H\to b\bar b$, $H\to c\bar c$, and $H\to gg$ (with gluon prongs);
  \item \textbf{Multi-prong hadronic decays:} three-prong top quarks
    ($t\to bqq'$) and four-prong Higgs decays ($H\to 4q$);
  \item \textbf{Semileptonic heavy-particle decays:} $t\to b\ell\nu$ and
    $H\to \ell\nu qq'$, containing a prompt electron or muon alongside hadronic
    activity.
\end{itemize}

We compare the internal representations with a reference set of observables,
grouped by IRC safety:
\begin{itemize}
  \item \textbf{IRC-safe jet observables:} ungroomed jet invariant mass
    $m_J = \sqrt{E_J^2 - |\vec{p}_J|^2}$; soft-drop groomed mass
    $\mSD$~\cite{Larkoski:2014wba}; jet width
    $w = (\sum_i p_{T,i}\Delta R_{i,J})/p_{T,J}$; unnormalised
    $N$-subjettiness $\tau_1,\tau_2,\tau_3,\tau_4$~\cite{Thaler:2010tr};
    and the 101 non-trivial normalised hadronic Energy Flow Polynomials
    (EFPs, angular exponent $\beta=1$, multigraph degree
    $d\le 5$)~\cite{Komiske:2017aww}, entered logarithmically as
    $\ln\max(\mathrm{EFP}, 10^{-12})$.
  \item \textbf{Sudakov-safe ratios:} the prongness ratios
    $\tau_{21}=\tau_2/\tau_1$ and $\tau_{32}=\tau_3/\tau_2$, which are not IRC
    safe but are calculable with resummation~\cite{Larkoski:2013paa,Larkoski:2015lea}.
  \item \textbf{IRC-unsafe jet observables:} constituent multiplicity
    $\nconst$ and transverse-momentum dispersion
    $\pTD = \sqrt{\sum_i p_{T,i}^2}/\sum_i p_{T,i}$~\cite{Gallicchio:2011xq,CMS:2013kfa},
    both sensitive to soft radiation and collinear splittings.
  \item \textbf{Per-particle observables (12 variables):} constituent momentum
    fraction $p_{T,i}/p_{T,J}$, $p_T$ rank within the jet, angular coordinates
    $(\Delta R_i, \Delta\eta_i, |\Delta\phi_i|)$, and (for interpretation only,
    withheld from \partkin's explicit feature vector) electric charge,
    transverse impact parameter $|d_0|$, and five binary particle-identification
    (PID) flags (charged hadron, neutral hadron, photon, electron, muon).
\end{itemize}

All evaluations use held-out jets that were not used to train SAEs or fit
erasers. Per-particle SAEs are trained on 80\% of jets
$0\text{--}3{,}999$ per file (200k jets total, $7.87\times 10^6$ particle tokens)
and evaluated on the remaining 20\% ($40{,}043$ held-out jets, fixed jet-level
split). The five-fold scaling test adds jets $4{,}000\text{--}19{,}999$ per file
($800\mathrm{k}$ extra jets, $960\mathrm{k}$ training jets total) only to SAE
training and keeps the same $40{,}043$ held-out jets. The
concept-erasure study fits erasers on jets $20{,}000\text{--}23{,}999$ per file
(200k jets) and evaluates on jets $24{,}000\text{--}24{,}999$ (50k jets). The
SAE group-ablation study uses jets $20{,}000\text{--}20{,}999$ (50k jets) to
compute decoded energies and nuisance correlations and evaluates on jets
$24{,}000\text{--}24{,}999$ (50k jets), both unseen by the SAEs though shared
with the erasure study. The IRC tests (Section~\ref{sec:irc}), the mass-removal test
(Section~\ref{sec:muons}) and the \partfull{} tests (Section~\ref{sec:full}) use
the same $50\mathrm{k}$ evaluation jets; the linear readouts of the IRC test and
the classifiers on the known observables are fitted on the $200\mathrm{k}$
erasure fit jets.

\subsection{Particle Transformer architecture and readout bottleneck}
\label{sec:arch}

\begin{figure*}[tbp]
\centering
\includegraphics[width=\textwidth]{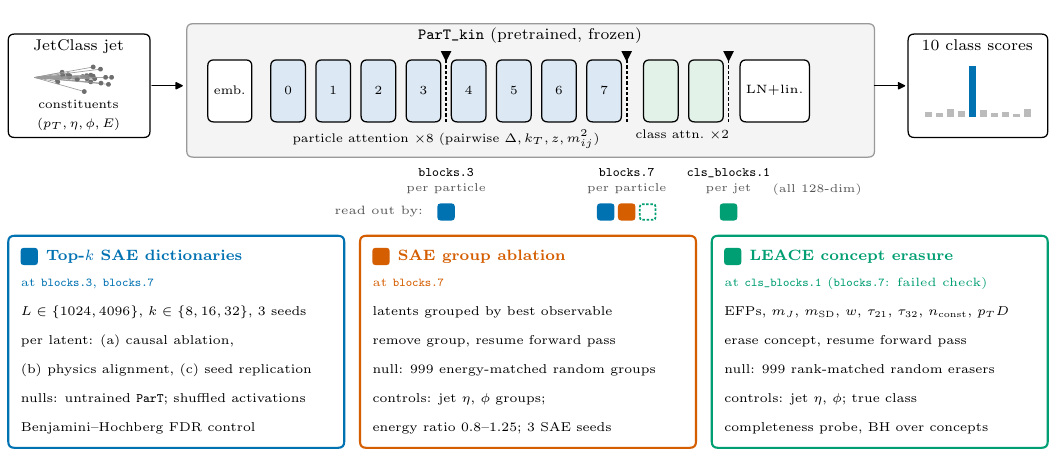}
\caption{\textbf{Methodology.} The frozen \partkin{} with its read-out points:
per-particle activations after particle-attention blocks 3 and 7 and the jet
vector after the last class-attention block (128 dimensions each). Coloured
squares under each read-out point mark the methods applied there. Top-$k$ SAE
dictionaries (blue) decompose the per-particle activations; each latent is tested
for a causal effect, alignment with known observables and replication across
seeds, against SAEs trained on an untrained \texttt{ParT} and on shuffled
activations, with false-discovery-rate control (Section~\ref{sec:sae_framework}).
Group ablation (orange) removes all SAE latents aligned with one observable at
\hook{blocks.7} and compares with energy-matched random groups; LEACE (green)
erases one concept from the jet vector and compares with rank-matched random
erasers; the token-wise version at \hook{blocks.7} (hollow square) failed its
azimuthal-symmetry check and is not used for results (Section~\ref{sec:concept}). In both, the forward pass resumes from the
edited activations and the accuracy loss is measured on held-out jets.}
\label{fig:method}
\end{figure*}

We study the public pretrained \partkin{} checkpoint~\cite{Qu:2022mxj} (\partfull{},
which adds PID flags and track impact parameters, validates our setup and is
used for the comparison of Section~\ref{sec:full}). For each constituent particle $i$, \partkin{} takes
seven standardised kinematic inputs---$\log p_{T,i}$, $\log E_i$,
$\log(p_{T,i}/p_{T,J})$, $\log(E_i/E_J)$, $\Delta R_i$, $\Delta\eta_i$, and
$\Delta\phi_i$---together with the constituent four-vectors
$p_i^\mu = (p_{x,i},p_{y,i},p_{z,i},E_i)$, from which a pairwise embedding
computes four Lorentz-motivated interaction variables ($\ln\Delta_{ij}$,
$\ln k_{T,ij}$, $\ln z_{ij}$, and $\ln m_{ij}^2$)~\cite{Qu:2022mxj} that bias
the self-attention logits. Eight particle-attention blocks
(\hook{blocks.0}--\hook{blocks.7}) evolve one 128-dimensional embedding per
constituent; two class-attention blocks (\hook{cls\_blocks.0}--\hook{cls\_blocks.1})
then pool the particle cloud into a single 128-dimensional jet token, followed
by a final LayerNorm (\hook{norm}) and a linear ten-class classifier head
(\hook{fc}). On the 50k held-out evaluation jets, \partkin{} attains an overall
accuracy of $0.7461$ and a macro one-vs-rest AUC of $0.9657$. We read out the
residual stream at three points (\emph{hooks}): halfway through the particle
blocks (\hook{blocks.3}), after the last particle block (\hook{blocks.7}), and
after the last class-attention block (\hook{cls\_blocks.1}). Figure~\ref{fig:method} summarises the model, the three
hooks and the three probes of Sections~\ref{sec:sae_framework}--\ref{sec:concept}, each with its null.

\paragraph{Low-dimensional readout.}
Only 22 of the 128 channels of the final LayerNorm (\hook{norm}) carry gain
$|\gamma|>10^{-2}$ in \partkin{}, and 17 principal components capture
$99.99\%$ of its variance over $10^6$ jets (106 in \hook{cls\_blocks.1}, 108 in
\hook{norm} of a randomly initialised \texttt{ParT}). SAEs therefore cannot
decompose \hook{norm} usefully, so we probe the residual stream upstream, and an
isotropic random direction in the 128-dimensional activations mostly misses the
subspace that reaches the classifier, which makes it a weak causal null
(Section~\ref{sec:criterion}).

\subsection{Two practical findings for JetClass and ParT users}
\label{sec:practical}

Two findings from the setup affect any study that reuses the public
\texttt{ParT} checkpoints or the JetClass kinematic inputs.

\paragraph{Silent framework incompatibility.}
The public 2022 \texttt{ParT} checkpoints load without warning into
\texttt{weaver-core}~0.5.3~\cite{weaver}, whose rewritten attention lowers the
accuracy on $3{,}000$ jets from $0.79$ to $0.27$. With version~0.4.17,
\partfull{} reaches $0.8595\pm 0.0020$ on $30{,}000$ validation jets, matching
the published $0.861$~\cite{Qu:2022mxj}; reusing these weights requires
\texttt{weaver-core}~0.4.x.

\paragraph{Kinematic inputs carry particle rest masses.}
In JetClass, Delphes four-vectors retain generator-level particle mass
hypotheses: charged hadrons carry the pion ($0.1396\,\mathrm{GeV}$, $81.8\%$),
kaon ($0.4937\,\mathrm{GeV}$, $12.1\%$), or proton ($0.9383\,\mathrm{GeV}$,
$6.2\%$) mass; muons carry $m_\mu = 0.1057\,\mathrm{GeV}$; and electrons,
photons, and neutral hadrons are effectively massless ($m \approx 0$). Although
\partkin{} receives no explicit PID flags, particle rest masses enter its
kinematic representation directly through two channels:
\begin{gather}
  \ln E_i - \ln p_{T,i}
  = \frac{1}{2}\ln\!\left(\cosh^2\!\eta_i + \frac{m_i^2}{p_{T,i}^2}\right),
  \label{eq:mass_leak}\\
  m_{ij}^2 = m_i^2 + m_j^2 + 2\left(E_i E_j - \vec{p}_i\cdot\vec{p}_j\right).
  \nonumber
\end{gather}
Because all constituents of a narrow jet share the jet pseudorapidity
$|\eta_J|$ up to the provided signed offset $\Delta\eta_i$ (and $\cosh^2\eta$
is even), comparing $\ln E_i - \ln p_{T,i}$ across constituents within a jet
isolates $m_i^2/p_{T,i}^2$, reinforced by the pairwise invariant masses
$m_{ij}^2$ and the rapidity--pseudorapidity shift $y_i - \eta_i$ inside
\texttt{ParT}'s pairwise attention bias. Section~\ref{sec:muons} shows
that \partkin{} uses these masses to build muon-specific features and that
removing them costs $0.106$ in accuracy. On JetClass,
``kinematics-only'' therefore means ``no explicit PID flags'', not
``PID-blind''. This also matters for IRC tests, where newly split or emitted particles must be assigned a consistent
mass hypothesis (Section~\ref{sec:irc}).

%==============================================================================
\section{Controlled sparse autoencoder framework}
\label{sec:sae_framework}
%==============================================================================

\subsection{Top-\texorpdfstring{$k$}{k} sparse autoencoders and why raw metrics fail}
\label{sec:sae_train}

To decompose activation vectors $x \in \mathbb{R}^{128}$ (centred and scaled to
unit mean squared norm), we train top-$k$ SAEs~\cite{Gao:2024sae}:
\begin{gather}
  z = \mathrm{TopK}\!\left(\mathrm{ReLU}\!\left(W_{\mathrm{enc}}(x - b_{\mathrm{dec}}) + b_{\mathrm{enc}}\right)\right),
  \label{eq:sae}\\
  \hat{x} = W_{\mathrm{dec}} z + b_{\mathrm{dec}},
  \nonumber
\end{gather}
with unit-norm decoder columns $d_j$ and sparsity enforced directly by $k$
rather than an $L_1$ shrinkage penalty. Adopting the training recipe of
ref.~\cite{Gao:2024sae} (decoder-parallel gradient projection, dead-latent
auxiliary loss, and geometric-median initialisation of $b_{\mathrm{dec}}$) cut
the fraction of held-out dead latents in per-particle SAEs from up to $72\%$ to
$2\text{--}41\%$. We sweep dictionary size $L\in\{1024, 4096\}$, sparsity
$k\in\{8, 16, 32\}$, and 3 random initialisations for 10 epochs at
\hook{blocks.3} and \hook{blocks.7}. Jet-level SAEs on \hook{cls\_blocks.1} kept
up to $63\%$ dead latents and replicated no better than the controls, so the jet
vector is probed with concept erasure instead (Section~\ref{sec:erasure}).

To test whether SAE features reflect \emph{learned physics} rather than
architecture or marginal statistics, every $(L,k)$ configuration is also trained
on two null controls:
\begin{enumerate}
  \item \textbf{Random-init control:} activations of an untrained \texttt{ParT}
    (same architecture, random weights, same jets), testing whether a feature
    requires training.
  \item \textbf{Shuffled control:} real \partkin{} activations with each of the
    128 coordinates permuted independently across tokens (separately within
    training and held-out splits), preserving every single-channel marginal
    distribution while destroying multi-channel geometry, the single-neuron
    baseline.
\end{enumerate}

As for language models~\cite{Heap:2025random}, standard SAE quality metrics do
not separate the trained model from these controls. On
held-out tokens, \partkin{} SAEs reach a fraction of variance unexplained (FVU)
of $0.006\text{--}0.065$ and cross-seed replication ($|\cos|>0.9$ with a live
partner in another seed) of $1\text{--}9\%$. Yet random-init SAEs achieve
\emph{comparable or lower} FVU ($0.006\text{--}0.037$) with $0.1\text{--}4.8\%$
replication, and shuffled-activation SAEs reach FVU $0.02\text{--}0.16$ and up
to $32\%$ replication (axis-aligned codes replicate readily). Uncontrolled SAE
diagnostics therefore cannot support physics claims.

\subsection{The calibrated learned-feature criterion}
\label{sec:criterion}

Before training any per-particle SAE, we fixed a three-part criterion for a
\emph{learned} feature, applied to the seed-0 latents active on at least
$10^{-4}$ of held-out tokens (the \emph{eligible} latents):
\begin{itemize}
  \item[(a)] \textbf{Causality.} On up to 256 held-out jets where latent $j$
    fires, we subtract its decoded contribution $s\, z_j d_j$ at the hook from
    every token where it fires; $s$ is the root-mean-square norm of the centred
    activations used to normalise the SAE input, so the subtracted vector is in
    \partkin{} units. We then finish the forward pass and measure the mean
    absolute logit shift $|\Delta\text{logit}|$, averaged over the ten classes
    and the jets.
  \item[(b)] \textbf{Physics alignment.} Each latent receives two held-out
    scores: its maximum absolute Spearman correlation $|\rho|$ across the 12
    per-particle observables (token level) and 11 jet observables (per-jet
    summed activation), and its 5-fold cross-validated ridge $R^2$ on the 101
    log EFPs ($d\le 5$). It must beat frequency-matched latents of \emph{both}
    controls.
  \item[(c)] \textbf{Cross-seed identifiability.} The latent must have a live
    partner with decoder $|\cos|>0.9$ in seed~1 or seed~2, and its $(L,k)$
    configuration's overall replication rate must exceed both controls'.
\end{itemize}

\paragraph{Why the isotropic null fails, and how the test is calibrated.}
In the initial pre-registered test, criterion~(a) compared each latent against
19 isotropic random directions carrying the same per-token coefficients. Applied
to ordinary latents, this passed $9\text{--}63\%$ of them instead of the nominal
$5\%$: isotropic directions are a weak null, probably because the
activations and the directions the network reads out are strongly anisotropic
(Section~\ref{sec:arch}). We
therefore amended the test, logging and validating the change on controls
before looking at real outputs (Appendix~\ref{app:protocol}): criterion~(a) now
takes the larger $p$-value, $p_a = \max(p_{\mathrm{cov}}, p_{\mathrm{shuf}})$, of
two nulls: (i) random directions drawn from the activation covariance,
$v \propto \Sigma^{1/2} g$, evaluated via Besag--Clifford sequential Monte
Carlo~\cite{Besag:1991mc}; and (ii) frequency-matched latents from disjoint
seeds of the shuffled-control SAEs. Criterion~(b) is expressed as a
Bonferroni-corrected empirical $p$-value $p_b$ against frequency-matched pools
of both controls. Within each hook and model, Benjamini--Hochberg (BH) at
$q=0.05$~\cite{Benjamini:1995fdr} controls the false discovery rate across all
eligible latents of the six $(L,k)$ configurations, with
Benjamini--Yekutieli (BY)~\cite{Benjamini:2001by} as a check against dependence
between latents. Because 499 null draws imposed $p$-value floors
($0.002$ for (a), $0.004$ for (b)) too coarse for BH across families of
$7.6\times 10^3\text{--}1.2\times 10^4$ latents, we raised resolution at
\hook{blocks.7} to $4{,}999$ draws and $4{,}999$-latent control pools (floors
$2\times 10^{-4}$ and $4\times 10^{-4}$; \hook{blocks.3} was not rerun because
even setting every floor $p$-value to zero yields at most 3 latents there).

\begin{table*}[t]
\centering
\caption{Calibration of the amended causal test on control latents: fraction
with $p_a \le 0.05$ under each null alone and under their maximum (nominal
$5\%$, pre-set acceptance ceiling $7.5\%$), plus BH discoveries at $q=0.05$.}
\label{tab:calibration}
\small
\begin{tabular}{llrrrrr}
\toprule
Draws & Control family & Latents & Null (i) cov. & Null (ii) shuf. & Combined & BH \\
\midrule
499     & shuffled, \hook{blocks.3}    & 1{,}200 (sample) & 9.5\% & 5.4\% & 1.1\% & 0 \\
499     & shuffled, \hook{blocks.7}    & 1{,}200 (sample) & 0.8\% & 6.7\% & 0.7\% & 0 \\
499     & random-init, \hook{blocks.3} & 1{,}200 (sample) & 7.4\% & 1.3\% & 0.1\% & 0 \\
499     & random-init, \hook{blocks.7} & 1{,}200 (sample) & 7.2\% & 2.2\% & 0.6\% & 0 \\
4{,}999 & shuffled, \hook{blocks.7}    & 9{,}564 (all)    & 0.5\% & 5.7\% & 0.4\% & 0 \\
4{,}999 & random-init, \hook{blocks.7} & 8{,}193 (all)    & 6.7\% & 1.8\% & 0.3\% & 0 \\
\bottomrule
\end{tabular}
\end{table*}

Table~\ref{tab:calibration} checks the calibration on control latents.
Neither null is calibrated alone (the covariance null passes up to $9.5\%$,
nearly twice the nominal rate), but their maximum is conservative
($0.1\text{--}1.1\%$ at $\alpha=0.05$) and gives no BH discoveries in any control
family. For physics alignment~(b), shuffled-control latents (evaluated on real
\partkin{} activations) beat the random-init control in $23\text{--}41\%$ of
cases, whereas random-init latents beat the shuffled control in only
$0.0\text{--}0.1\%$: any dictionary applied to real activations inherits their
single-channel physics marginals, so the shuffled control is the stricter
alignment null.

%==============================================================================
\section{SAE results: population physics without single-latent identifiability}
\label{sec:sae_results}
%==============================================================================

\subsection{Population-level physics enrichment and layer progression}
\label{sec:population}

Table~\ref{tab:population} compares the
fraction of eligible seed-0 latents passing each criterion nominally
($p\le 0.05$) in trained \partkin{} with both controls under the same rule, with Newcombe hybrid-score $95\%$ confidence
intervals~\cite{Newcombe:1998ci} and two-sided Fisher exact $p$-values. Because
latents within an SAE share training data and null pools, pooled Fisher
$p$-values overstate statistical independence; consistency across the six
$(L,k)$ configurations provides the more robust test.

\begin{table*}[t]
\centering
\caption{Population-level pass rates ($p \le 0.05$) of eligible seed-0 latents
in trained \partkin{} (Kin) versus each control under the same rule. Differences
are in percentage points with Newcombe $95\%$ CIs; Fisher $p$-values compare Kin
vs.\ Shuffled. \hook{blocks.7} uses $4{,}999$ draws over full families
($12{,}211$ Kin, $9{,}564$ Shuffled, $8{,}193$ Random-init); \hook{blocks.3}
uses $499$ draws ($7{,}615$ Kin, $1{,}200$ sampled latents per control).}
\label{tab:population}
\fitwidth{%
\begin{tabular}{llrrlrl}
\toprule
Hook & Criterion & Kin & Shuffled & Diff.\ (95\% CI), Fisher $p$ & Random-init & Diff.\ (95\% CI) \\
\midrule
\multirow{3}{*}{\hook{blocks.7}}
 & (a) Causal      &  7.5\% & 0.37\% & $+7.1$ (6.7, 7.6), $10^{-185}$           & 0.27\% & $+7.2$ (6.8, 7.7) \\
 & (b) Alignment   & 14.8\% & 4.1\%  & $+10.7$ (10.0, 11.5), $4\times 10^{-164}$ & 0.05\% & $+14.7$ (14.1, 15.4) \\
 & (c) Replication &  4.5\% & 3.8\%  & $+0.7$ (0.2, 1.3), $0.007$               & 3.0\%  & $+1.5$ (1.0, 2.0) \\
\midrule
\multirow{3}{*}{\hook{blocks.3}}
 & (a) Causal      &  8.3\% & 1.1\%  & $+7.2$ (6.2, 8.0), $2\times 10^{-26}$    & 0.08\% & $+8.2$ (7.5, 8.8) \\
 & (b) Alignment   &  4.8\% & 2.4\%  & $+2.3$ (1.2, 3.2), $10^{-4}$             & 0\%    & $+4.8$ (4.2, 5.3) \\
 & (c) Replication &  6.0\% & 21.9\% & $-16.0$ ($-18.4$, $-13.6$)               & 3.6\%  & $+2.4$ (1.1, 3.4) \\
\bottomrule
\end{tabular}}
\end{table*}

Two results stand out at the population level:
\begin{enumerate}
  \item \textbf{Enrichment over the controls:} At \hook{blocks.7}, $7.5\%$
    of trained latents pass the calibrated causal test (vs.\ $0.37\%$ shuffled
    and $0.27\%$ random-init, exceeding both controls in 5 of 6 configurations),
    and $14.8\%$ pass physics alignment (vs.\ $4.1\%$ shuffled and $0.05\%$
    random-init, beating the shuffled control in 4 of 6 configurations).
  \item \textbf{Depth dependence (two layers):} the causal excess is already
    present halfway through the particle-attention stack at \hook{blocks.3}
    ($8.3\%$ vs.\ $1.1\%$ and $0.08\%$), whereas the alignment excess is
    larger at \hook{blocks.7} ($14.8\%$, $+10.7$ points over shuffled) than at
    \hook{blocks.3} ($4.8\%$, $+2.3$ points). With only two layers, tested at
    different Monte Carlo resolutions and with sampled versus full control
    families, this suggests, but does not measure, a progressive assembly of
    jet-substructure variables across the particle-attention stack.
\end{enumerate}
Cross-seed replication~(c), by contrast, shows no consistent excess over the
controls: at \hook{blocks.3} the shuffled control replicates $21.9\%$ of latents
against $6.0\%$ for \partkin{}, and even at \hook{blocks.7} the shuffled control
replicates $74\%$ of latents at $L=1024, k=32$.

\subsection{Individual feature certification fails, and replication binds}
\label{sec:perfeature}

Under the uncalibrated pre-registered test (isotropic causal null), 58
latent--configuration pairs (54 distinct decoder directions) passed all three
criteria, mostly aligned with jet mass and soft-drop mass (20), muon identity
(18), $\pTD$, $N$-subjettiness ratios, jet width, and multiplicity. Because that
causal null passed $9\text{--}63\%$ of ordinary latents, those 58 pairs are
diagnostic counts, not certified claims.

\begin{table}[t]
\centering
\caption{Cut-flow of individual seed-0 SAE latents at \hook{blocks.7}
($4{,}999$ null draws, six $(L,k)$ configurations pooled, $40{,}043$ held-out
jets). Nominal stages use $p\le 0.05$; the last row applies Benjamini--Hochberg
at $q=0.05$ over all eligible latents. For comparison, the causal test passes
$0.37\%$ of shuffled-control latents, i.e.\ about 45 of $12{,}211$.}
\label{tab:cutflow}
\begin{tabular}{lr}
\toprule
Stage & Latents \\
\midrule
Eligible (active on $\ge 10^{-4}$ of tokens) & $12{,}211$ \\
(a) causal, nominal & 917 \\
\quad and (b) physics-aligned, nominal & 259 \\
\quad and (c) seed partner, $|\cos|>0.9$ & 37 \\
\quad and configuration gate & 16 \\
All three after BH & \textbf{0} \\
\bottomrule
\end{tabular}
\end{table}

Under the calibrated criterion with BH correction ($q=0.05$), \textbf{zero
individual latents pass all three tests} (Table~\ref{tab:cutflow}) at either hook, at either $499$ or
$4{,}999$ null draws, and BH rejects nothing for (a) or (b) individually (BY
likewise). At \hook{blocks.7} with $4{,}999$ draws, this null result is not a
Monte Carlo resolution artefact: 21 latents sit at the causal
floor (where BH requires 49) and 93 sit at the alignment floor (98 required).
Even with infinite Monte Carlo resolution (every floor $p$-value set to zero),
at most \textbf{one} latent would pass all three criteria, a muon-specific feature ($L=4096, k=16$, latent 3945;
Section~\ref{sec:muons}).

\paragraph{Replication binds, also with $5\times$ more data.}
Among trained latents at \hook{blocks.7} that pass the other two tests
nominally, cross-seed replication~(c) eliminates $86\%$ ($222/259$), against
$72\%$ for alignment and $56\%$ for causality; even causal-and-aligned latents,
which replicate three times more often than typical ones ($14.3\%$ vs.\
$4.5\%$), mostly fail to reappear in another seed. Retraining the \hook{blocks.7}
SAEs on $10^6$ jets ($960\mathrm{k}$ for training, the same $40{,}043$ held-out
jets, matched optimiser steps) barely moves the configuration-level replication
rates ($0.087\to0.099$ at best; $1.6\text{--}9.9\%$ overall), and the same three
configurations beat both controls ($L/k = 1024/8$, $4096/16$, $4096/32$). Again
no latent is certified. With $10^6$ jets, 98 latents pass BH for alignment~(b)
alone, but \emph{all of them fail the causal test~(a)}, while the population
excess is unchanged (causal $8.2\%$ vs.\ $0.26\%$ shuffled; alignment $14.9\%$
vs.\ $3.9\%$).

\paragraph{A possible geometric picture.}
Unlike discrete lexical tokens in language models, collider jet observables ($m_J$, $\mSD$, $w$, $\tau_N$, EFPs)
are continuous and strongly correlated, and may be represented redundantly
across constituent tokens, or in superposition~\cite{Elhage:2022toy}. Different
SAE seeds could then tile the same physics subspace with rotated or split
dictionaries of similar quality, so that no single atom is canonical, as also
reported for SAEs on language models~\cite{Paulo:2025same}; consistent with this, single-atom ablations have small
effects (mean $|\Delta\text{logit}|$ of $0.007\text{--}0.87$ for the
pre-registered passes). The cross-seed group ablation of
Section~\ref{sec:groupablation} supports this picture: groups aligned with the
same observable replicate in effect and partly overlap in subspace across
seeds, while sharing few individual latents.

\subsection{Hidden particle-mass cues: muon-specific features and reliance on masses}
\label{sec:muons}

Before turning to jet-level concepts, we examine one unexpected group of
aligned per-particle latents: those whose best match is muon identity, although
\partkin{} receives no PID flags. Could these latents merely detect hard,
isolated particles (a kinematic signature of prompt leptons in $t\to b\ell\nu$
and $H\to\ell\nu qq'$)? To separate kinematics from particle rest mass, we
reweighted every particle type (electrons, charged hadrons, neutral hadrons,
photons) to match the muons' joint distribution in $p_T$ rank ($0$, $1\text{--}2$,
$3\text{--}9$, $\ge 10$) and relative isolation ($\sum_{j\neq i,\Delta R<0.4}
p_{T,j}/p_{T,i}$ in bins $[0,0.1,0.5,2,\infty)$), calling a latent
\emph{muon-specific} if its kinematics-matched firing rate on muons is at least
three times that on electrons and on every hadron/photon species.

Of all seed-0 latents whose best-matching observable is muon identity, $51/64$
at \hook{blocks.3} and $94/112$ at \hook{blocks.7} are muon-specific,
firing on up to $\sim 60\%$ of muons and rarely on kinematically matched
electrons, hadrons, or photons. Seventeen of the 18 pre-registered muon pairs
are muon-specific, as is the sole infinite-resolution candidate at
\hook{blocks.7} (matched firing rates: $33\%$ on muons, $10\%$ on electrons,
$5\%$ on charged hadrons). This is consistent with the mechanism of
eq.~\eqref{eq:mass_leak}, in which \partkin{} reads the muon rest mass
($0.1057\,\mathrm{GeV}$, distinct from $m_e \approx 0$ and
$m_{\pi,K,p} \ge 0.1396\,\mathrm{GeV}$) encoded in the JetClass four-vectors;
the reweighting controls $p_T$ rank and isolation, but other differences between
simulated muons and electrons are not excluded by this test. It offers one
possible explanation, for muons though not for massless electrons, of the
observation in ref.~\cite{Legge:2025cnm} that a kinematic \texttt{ParT} attends
strongly to the lepton in semileptonic top decays.

\paragraph{Removing the masses.}
To measure how much the output depends on these masses, we changed the
constituent energies of the $50{,}000$ evaluation jets to
$E_i=(|\vec p_i|^2+m_i^2)^{1/2}$ with a new mass, keeping every three-momentum,
in a pre-registered test with three variants and one control
(Table~\ref{tab:massremoval}; paired bootstrap over jets, $1{,}000$ resamples,
per-class recall changes claimed at BH $q=0.05$ over the 10 classes, which
holds for at least 9 classes in every variant). Setting
every mass to zero costs $0.106$; the muon mass alone costs $0.016$, with the
largest loss in $H\to b\bar b$ ($-0.099$ in recall); giving all charged hadrons
the pion mass, which removes kaon and proton identification, costs $0.022$,
again largest in $H\to b\bar b$ ($-0.104$). A plausible reading, not tested
here, is that \partkin{} uses the masses to recognise non-isolated muons and
kaons from $b$-hadron decays. The registered expectation that the semileptonic
classes would lose most did not hold: although jets with a muon are most common
in $t\to b\ell\nu$ ($58.5\%$) and $H\to\ell\nu qq'$ ($52.3\%$), their prompt
leptons are hard and isolated, and these classes lose less ($-0.017$ and
$-0.037$ for massless muons). The control, which permutes the masses among the
constituents of each jet, is as damaging as removing them ($-0.115$), so it is
not a null: it shows that which particle carries which mass matters. All these
inputs lie outside the training distribution, so the drops measure the
dependence of \partkin{} on the generator masses, not the accuracy a tagger
trained without them would lose.

\begin{table}[t]
\centering
\caption{Removing generator particle masses from the \partkin{} inputs
(energies only; three-momenta kept), with respect to the unmodified jets
(accuracy $0.746$, macro AUC $0.966$; $50{,}000$ jets). All 95\%
paired-bootstrap intervals of $\Delta$Acc have half-widths $\le0.004$ and exclude
zero. Flips: jets whose predicted class changes. Last column: largest
per-class recall loss.}
\label{tab:massremoval}
\begin{tabular}{lcccl}
\toprule
Inputs changed & $\Delta$Acc & $\Delta$AUC & Flips & Largest loss \\
\midrule
All massless       & $-0.106$ & $-0.023$ & $27\%$ & $H\to gg$, $-0.43$ \\
Muons massless     & $-0.016$ & $-0.004$ & $4\%$  & $H\to b\bar b$, $-0.10$ \\
$K, p\to m_\pi$    & $-0.022$ & $-0.006$ & $10\%$ & $H\to b\bar b$, $-0.10$ \\
Masses permuted    & $-0.115$ & $-0.038$ & $25\%$ & $H\to\ell\nu qq'$, $-0.27$ \\
\bottomrule
\end{tabular}
\end{table}

%==============================================================================
\section{Concept-level causal analysis: what jet physics \texorpdfstring{\partkin{}}{ParT\_kin} relies on}
\label{sec:concept}
%==============================================================================

Because individual SAE latents are not identifiable across seeds and
single-latent ablations have small effects, we ask the causal question of
\emph{whole physics concepts}, with two complementary probes:
dictionary-free linear concept erasure (LEACE) at \hook{cls\_blocks.1} and
\hook{blocks.7} (Section~\ref{sec:erasure}), and energy-matched group ablation
of physics-aligned SAE latents at \hook{blocks.7}
(Section~\ref{sec:groupablation}).

\subsection{Linear concept erasure (LEACE)}
\label{sec:erasure}

LEACE~\cite{Belrose:2023leace} removes all linearly extractable information
about a target concept $Z$ from activations $X$ via the closed-form oblique
projection
\begin{equation}
  r(x) = x - W^+ P W (x - \mu),
  \label{eq:leace}
\end{equation}
where $\mu = \mathbb{E}[X]$, $W = \Sigma_{xx}^{-1/2}$ is the symmetric whitener,
$W^+$ its pseudoinverse, and $P$ is the orthogonal projection onto the column
space of $W\Sigma_{xz}$ (dropping singular values $\le 0.01$ after standardising
concept columns). By construction $\mathrm{Cov}(r(X), Z) = 0$ with minimal $L_2$
distortion, ensuring no linear probe beats a constant predictor on the fit
distribution.

To erase non-monotonic dependencies, such as a resonant mass window around
$m_W$, $m_Z$, or $m_H$, each of the seven named scalar observables ($m_J$,
$\mSD$, $\tau_{21}$, $\tau_{32}$, $\nconst$, $\pTD$, $w$) is encoded as one-hot
indicators of 20 equal-population quantile bins. In addition, the 101 log EFPs
($d\le 5$) are erased as a continuous block, as is the union of all known
observables (\texttt{known\_all}: log EFPs plus the seven unbinned continuous
scalars). Erasers are fitted on $200\mathrm{k}$ jets and evaluated on
$50\mathrm{k}$ held-out jets.
Each concept's held-out accuracy loss $\Delta\mathrm{Acc}$ is benchmarked
against $999$ rank-matched random subspaces drawn in the same whitened
coordinates from the between-jet covariance $W B W$ (matching the fact that
jet-level concepts correlate only with between-jet variation at token hooks).

One positive control, two negative controls and a completeness gate govern every claim:
\begin{itemize}
  \item \textbf{Positive control:} erasing the true ten-class label must
    collapse classification accuracy.
  \item \textbf{Negative controls:} \partkin's inputs are invariant under
    global azimuthal rotations (jet $\phi$), and jet $\eta$ enters only through
    the particle energies and is not expected to matter for classification. If either
    jet $\phi$ or jet $\eta$ has $p \le 0.05$, the hook's null is declared
    miscalibrated and \emph{no} claims are made at that hook.
  \item \textbf{Erasure completeness gate:} an erasure is accepted as complete
    only if a 2-fold held-out linear probe after erasure falls to at most
    $\text{chance} + 0.02$ (categorical) or $R^2 \le 0.02$ (continuous). Claims
    require passing both BH ($q=0.05$ over the nine physics concepts) and
    completeness at a calibrated hook.
\end{itemize}
Because LEACE leaves non-linear functions of $Z$ intact, the accuracy remaining
after erasing \texttt{known\_all} is an \emph{upper bound} on what the
model uses beyond known observables. At \hook{blocks.7}, where erasure acts
token-by-token upstream of two non-linear class-attention layers, accuracy
losses are \emph{lower bounds} on reliance.

\begin{table*}[t]
\centering
\caption{Linear concept erasure (LEACE) in \partkin{} on $50\mathrm{k}$ held-out
jets (base accuracy $0.746$, macro AUC $0.966$): erased rank, accuracy loss
$\Delta\mathrm{Acc}$, mean loss of $999$ matched null erasers, one-sided
$p$-value (floor $0.001$), and post-erasure linear probe versus completeness
limit (2-fold accuracy for binned concepts, $R^2$ for continuous blocks).
\hook{blocks.7} fails the jet-$\phi$ negative control, so no claims are made
there.}
\label{tab:erasure}
\fitwidth{%
\begin{tabular}{llrrrrllrrrrl}
\toprule
& & \multicolumn{6}{c}{\hook{cls\_blocks.1} (calibrated)} & \multicolumn{5}{c}{\hook{blocks.7} (not calibrated, no claims)} \\
\cmidrule(lr){3-8}\cmidrule(lr){9-13}
Concept & Type & Rank & Loss & Null & $p$ & Probe / limit & Status & Rank & Loss & Null & $p$ & Probe / limit \\
\midrule
All known obs. & Combined   & 17 & 0.333 & 0.047 & 0.001 & $-$0.002 / 0.020 & \textbf{claimed} & 15 & 0.145 & 0.044 & 0.001 & $-$0.004 / 0.020 \\
EFPs ($d\le 5$) & IRC-safe  & 11 & 0.252 & 0.028 & 0.001 & $-$0.002 / 0.020 & \textbf{claimed} & 10 & 0.104 & 0.029 & 0.001 & $-$0.004 / 0.020 \\
Jet mass $m_J$  & IRC-safe  & 19 & 0.268 & 0.053 & 0.001 & 0.099 / 0.072    & incomplete       & 14 & 0.163 & 0.041 & 0.001 & 0.067 / 0.093 \\
Soft-drop $\mSD$& IRC-safe  & 19 & 0.243 & 0.053 & 0.001 & 0.100 / 0.072    & incomplete       & 16 & 0.150 & 0.047 & 0.001 & 0.066 / 0.088 \\
Jet width $w$   & IRC-safe  & 19 & 0.213 & 0.053 & 0.001 & 0.079 / 0.072    & incomplete       & 13 & 0.137 & 0.038 & 0.001 & 0.065 / 0.091 \\
$\tau_{21}$     & Sudakov-safe & 19 & 0.170 & 0.054 & 0.001 & 0.065 / 0.071    & \textbf{claimed} & 14 & 0.099 & 0.041 & 0.001 & 0.060 / 0.080 \\
$\tau_{32}$     & Sudakov-safe & 19 & 0.123 & 0.054 & 0.001 & 0.063 / 0.072    & \textbf{claimed} & 12 & 0.072 & 0.035 & 0.001 & 0.057 / 0.074 \\
Multiplicity    & IRC-unsafe& 19 & 0.144 & 0.053 & 0.001 & 0.079 / 0.084    & \textbf{claimed} & 12 & 0.067 & 0.035 & 0.002 & 0.093 / 0.119 \\
$\pTD$          & IRC-unsafe& 19 & 0.123 & 0.054 & 0.001 & 0.077 / 0.072    & incomplete       & 13 & 0.049 & 0.038 & 0.139 & 0.082 / 0.109 \\
\midrule
True class      & Pos.\ ctrl &  9 & 0.652 & 0.022 & 0.001 & 0.173 / --       & control          &  9 & 0.197 & 0.025 & 0.001 & 0.151 / -- \\
Jet $\eta$      & Neg.\ ctrl & 19 & 0.041 & 0.053 & 0.945 & 0.050 / --       & control          & 13 & 0.025 & 0.038 & 0.928 & 0.054 / -- \\
Jet $\phi$      & Neg.\ ctrl & 19 & 0.063 & 0.053 & 0.122 & 0.049 / --       & control          & 12 & 0.070 & 0.035 & 0.001 & 0.053 / -- \\
\bottomrule
\end{tabular}}
\end{table*}

Table~\ref{tab:erasure} presents the results.

\paragraph{Calibration outcomes.}
At \hook{cls\_blocks.1} both checks pass: erasing the true class reduces
accuracy from $0.746$ to $0.094$ (chance $=0.10$), and both negative controls
are consistent with the null (jet $\eta$: $p=0.945$; jet $\phi$: $p=0.122$). At
\hook{blocks.7}, erasing jet $\phi$ costs $0.070$, above all $999$ null draws
($p=0.001$), although a linear probe finds no azimuth information in the
activations; the cause is not established. By the pre-registered rule, no LEACE
claims are made at \hook{blocks.7}.

\paragraph{Claims at \hook{cls\_blocks.1}.}
BH at $q=0.05$ rejects all nine physics concepts at $p=0.001$, and the
completeness gate leaves \textbf{five} claims (unchanged under the alternative
BH reading; Appendix~\ref{app:protocol}). Erasing all known observables
together (rank 17) costs $\Delta\mathrm{Acc}=0.333$ (null $0.047$), cutting
accuracy from $0.746$ to $0.413$ (macro AUC $0.966\to0.834$). Erasing the EFP
basis (rank 11 of 128) costs $0.252$ (null $0.028$), the Sudakov-safe ratios
$\tau_{21}$ and $\tau_{32}$ cost $0.170$ and $0.123$ (null $0.054$), and the
IRC-unsafe multiplicity costs $0.144$ (null $0.053$). Erasing $m_J$, $\mSD$, $w$
or $\pTD$ also costs far more than the null ($0.123\text{--}0.268$), but a probe
still finds these binned concepts after erasure (accuracy $0.077\text{--}0.100$
against a limit of $0.072$), so they are not claimed from LEACE.

\paragraph{Where the remaining accuracy lies (exploratory).}
After erasing all known observables, the recall losses (exploratory,
uncorrected) are large for $t\to bqq'$ ($0.83$), the two semileptonic classes,
$H\to 4q$, QCD and $H\to gg$ ($0.37\text{--}0.58$), but compatible with the null
for the two-prong $W$, $Z$, $H\to b\bar b$ and $H\to c\bar c$ jets ($p\ge
0.19$). One reading is that linear erasure of the unbinned scalars leaves
non-monotonic information such as a mass window around $m_W$, $m_Z$ or $m_H$,
but per-class recall also moves when predictions shift between classes, so the
pattern does not test it.

\paragraph{What the known observables achieve alone.}
The $0.413$ left after erasure is an upper bound (Section~\ref{sec:erasure}).
As a counterpart we trained three pre-registered classifiers on the same
108 observables (101 log EFPs and the seven scalars) with the same $200\mathrm{k}$
fit and $50\mathrm{k}$ evaluation jets: a multilayer perceptron reaches an
accuracy of $0.600$ [$0.595$, $0.604$] (macro AUC $0.923$), gradient-boosted
trees $0.578$ and logistic regression $0.537$, against $0.746$ for \partkin{}
on the same jets. The perceptron's accuracy still rises between $100\mathrm{k}$
and $200\mathrm{k}$ fit jets ($+0.007$ [$+0.004$, $+0.009$]), so by the
registered reading the gap of $0.146$ [$0.142$, $0.150$] is an upper estimate of
what \partkin{} uses beyond these observables, not a measurement. The largest
per-class shortfalls are in $H\to b\bar b$ ($-0.30$ in recall), $H\to c\bar c$
($-0.22$) and $t\to b\ell\nu$ ($-0.21$), where flavour and the lepton carry
information that these observables lack and that \partkin{} can partly read
from the generator masses (Section~\ref{sec:muons}).

\subsection{Group ablation of physics-aligned SAE latents at \texorpdfstring{\hook{blocks.7}}{blocks.7}}
\label{sec:groupablation}

As a complementary test on the SAE representation at \hook{blocks.7}, where
token-wise LEACE failed its calibration, we ablate whole groups of
observable-aligned SAE latents at once; this also tests jet mass and $\pTD$,
whose linear erasure at \hook{cls\_blocks.1} was incomplete. We use the
seed-0 SAEs of the three configurations passing the replication gate of
criterion~(c): $(L,k)=(1024,8)$, $(4096,16)$, and $(4096,32)$. Eligible latents
with best $|\rho_{\mathrm{Spearman}}| \ge 0.3$ (labelled on the original
$40\mathrm{k}$ held-out split) are grouped by their best-matching observable,
testing every group with at least 10 latents.

Ablating a group subtracts the decoded contribution of all member latents from
every particle token where they fire at \hook{blocks.7} (retaining the SAE
reconstruction residual) and finishes the forward pass on the $50\mathrm{k}$
evaluation jets. The null is $999$ random groups of the same size from the same SAE, each
member matched in decoded activation energy (on disjoint $50\mathrm{k}$
matching jets), since larger perturbations cost more accuracy. Four
pre-registered rules govern claims per configuration: (1)~ablating every latent
must cost at least $0.10$; (2)~negative-control groups of unaligned latents
most correlated with jet $\phi$ and $\eta$ must have $p>0.05$; (3)~a group's
decoded energy must be $0.8\text{--}1.25$ times its nulls'; (4)~BH at $q=0.05$
over the physics groups of the configuration. An exploratory second null
($p_{\mathrm{al}}$), drawn only from latents aligned with other observables,
tests whether an effect is specific to its observable.

\begin{table*}[t]
\centering
\caption{Cross-configuration summary of SAE group ablation at \hook{blocks.7}, SAE seed~0
($50\mathrm{k}$ held-out jets, unmodified accuracy $0.7461$; ablating all
latents leaves $0.1871$, $0.1971$, and $0.1769$). Entries show accuracy loss
(null mean) and status: \textbf{bold} = claimed; $^\dagger$ = energy ratio
outside $0.8\text{--}1.25$ (not claimed); $^*$ = claimed but not separated from
other physics-aligned groups under the exploratory aligned-only null
($p_{\mathrm{al}}=0.127$); ``--'' = fewer than 10 latents.
Complete per-configuration tables including group sizes, $p$, energy ratios,
and $p_{\mathrm{al}}$ are given in the Supplemental Material~\cite{supp}. Control $p$-values: jet $\phi$
$0.449$, $0.257$, $0.404$; jet $\eta$ $0.808$, $0.934$, $0.980$ (columns left to right).}
\label{tab:groupabl_summary}
\fitwidth{%
\begin{tabular}{llccc}
\toprule
Observable group & Physics type & $L=1024,\ k=8$ & $L=4096,\ k=16$ & $L=4096,\ k=32$ \\
\midrule
Jet mass $m_J$   & IRC-safe   & \textbf{0.0808} (0.0131) & \textbf{0.0466} (0.0081) & \textbf{0.0644} (0.0135) \\
$\tau_2$         & IRC-safe   & \textbf{0.0079} (0.0034) & \textbf{0.0036} (0.0004) & \textbf{0.0096} (0.0010) \\
$\pTD$           & IRC-unsafe & 0.0090 (0.0033)          & \textbf{0.0261} (0.0023) & \textbf{0.0595} (0.0039) \\
Jet width $w$    & IRC-safe   & \textbf{0.0262} (0.0024) & 0.0000 (0.0014)          & $-$0.0002 (0.0000) \\
$\tau_3$         & IRC-safe   & \textbf{0.0088} (0.0016) & 0.0036 (0.0015)          & 0.0030 (0.0021) \\
Soft-drop $\mSD$ & IRC-safe   & \textbf{0.0078} (0.0013) & 0.0004 (0.0004)          & 0.0011 (0.0009) \\
$\tau_{21}$      & Sudakov-safe & \textbf{0.0036}$^*$ (0.0011) & 0.0011 (0.0005)      & 0.0017 (0.0022) \\
$\tau_1$         & IRC-safe   & 0.0006 (0.0013)          & 0.0362$^\dagger$ (0.0030)& 0.0109 (0.0036) \\
$\tau_4$         & IRC-safe   & 0.0034$^\dagger$ (0.0081)& 0.0021 (0.0092)          & 0.0077 (0.0147) \\
Multiplicity     & IRC-unsafe & 0.0062$^\dagger$ (0.0201)& 0.0028 (0.0300)          & 0.0139$^\dagger$ (0.0881) \\
Muon identity    & Hidden PID & --                       & 0.0011 (0.0005)          & 0.0008 (0.0008) \\
\midrule
Jet $\phi$ (ctrl)& Symmetry   & 0.0003 (0.0002) & 0.0001 ($-$0.0001) & 0.0000 ($-$0.0001) \\
Jet $\eta$ (ctrl)& Symmetry   & 0.0001 (0.0006) & $-$0.0003 (0.0001) & $-$0.0004 (0.0010) \\
\bottomrule
\end{tabular}}
\end{table*}

Table~\ref{tab:groupabl_summary} summarises the
outcomes (full per-configuration tables are in the Supplemental Material~\cite{supp}).

\paragraph{Calibration and claims.}
All three configurations pass rules~1 and~2: ablating every latent drops
accuracy from $0.746$ to $0.187$, $0.197$, and $0.177$, and all six symmetry
controls are consistent with their nulls (jet $\phi$: $p=0.449, 0.257, 0.404$;
jet $\eta$: $p=0.808, 0.934, 0.980$). Jet mass and $\tau_2$ are claimed in \textbf{all three} configurations. Jet
mass has the largest effect in every dictionary ($\Delta\mathrm{Acc} = 0.081,
0.047, 0.064$ against null means $0.013, 0.008, 0.014$), $\tau_2$ smaller ones
($0.004\text{--}0.010$); since unnormalised $\tau_2$ correlates with jet mass,
the two groups may remove overlapping physics.
In addition, IRC-unsafe \textbf{$\pTD$} is claimed in both $L=4096$
configurations (in seed~0 only; it does not replicate across seeds, see below) with substantial losses ($0.0261$ and $0.0595$ vs.\ nulls
$0.0023$ and $0.0039$; $p=0.085$ at $L=1024,k=8$), while jet width ($0.0262$),
$\tau_3$ ($0.0088$), $\mSD$ ($0.0078$), and $\tau_{21}$ ($0.0036$) are claimed
only at $L=1024,k=8$. The energy-matching rule catches one confound: $\tau_1$
at $L=4096,k=16$ yields $\Delta\mathrm{Acc}=0.0362$ ($p=0.001$) but carries
$2.01\times$ the decoded energy of its nulls, so it is flagged and not claimed.
Against the exploratory aligned-only null, \textbf{11 of the 12 claimed
groups} remain separated ($p_{\mathrm{al}}\le 0.006$, energy ratios
$0.83\text{--}1.07$; uncorrected), which suggests that the effects are specific
to the observable; the exception
is $\tau_{21}$ at $L=1024,k=8$ ($p_{\mathrm{al}}=0.127$).

\paragraph{Interpretation and contrast with LEACE.}
A group-ablation claim establishes that a set of latents aligned with an
observable matters causally more than equal-energy random latents in that SAE;
it does not show that the network computes that observable, and
because the SAE error term remains and class attention can recombine unablated
tokens, losses at \hook{blocks.7} are lower bounds on reliance. All
configurations are trained on the same activations, so agreement across
configurations is descriptive; robustness to the SAE initialisation is tested
next.

\begin{table*}[t]
\centering
\caption{Group ablation across SAE seeds at \hook{blocks.7}: the seeds (0, 1,
2) in which each observable group is claimed under rules~1--4, per
configuration. \textbf{Bold}: replicated (claimed in at least two of three
seeds, the pre-registered cross-seed rule). $^\dagger$: rejected by BH but
outside the energy window of rule~3, so not claimed. Groups never claimed in
any seed (including $\nconst$, $\tau_1$, $\tau_4$ and muon identity) are
omitted. Rules~1 and~2 pass in all nine seed--configuration pairs.}
\label{tab:groupseeds}
\begin{tabular}{lccc}
\toprule
Observable group & $L=1024,\ k=8$ & $L=4096,\ k=16$ & $L=4096,\ k=32$ \\
\midrule
Jet mass $m_J$   & \textbf{0, 1, 2} & \textbf{0, 1, 2} & \textbf{0, 1, 2} \\
$\tau_2$         & 0                & \textbf{0, 2}    & \textbf{0, 1, 2} \\
$\tau_3$         & \textbf{0, 1}    & 1                & 2 \\
Soft-drop $\mSD$ & \textbf{0, 1}    & --               & -- \\
Jet width $w$    & \textbf{0, 2}    & --               & -- \\
$\tau_{21}$      & 0                & --               & -- \\
$\pTD$           & (2$^\dagger$)    & 0                & 0 (2$^\dagger$) \\
\bottomrule
\end{tabular}
\end{table*}

\paragraph{Replication across SAE seeds.}
Individual latents do not replicate across SAE seeds
(Section~\ref{sec:perfeature}), so we repeated the group ablation for seeds~1
and~2 of the same three configurations, pre-registered before any new label or
number existed: each seed's latents were labelled as for seed~0, and the design
and rules were left unchanged. A group counts as replicated in a configuration
if it is claimed in at least two of the three seeds
(Table~\ref{tab:groupseeds}). All new seed--configuration pairs pass rules~1
and~2. Jet mass is claimed in every seed of every configuration (accuracy loss
$0.039\text{--}0.091$ against null means $0.005\text{--}0.017$), and $\tau_2$
replicates in both $L=4096$ configurations. Of the twelve seed-0 claims, eight
replicate. The four that do not are $\tau_2$ and $\tau_{21}$ at $L=1024,k=8$
and $\pTD$ in both $L=4096$ configurations: in seeds~1 and~2 its groups fall
outside the energy window of rule~3 (ratios $0.71\text{--}1.49$), and three of
these four are not rejected by BH either (only seed~2 at $L=4096,k=32$ is, at
energy ratio $0.79$). Among the replicated groups, sizes vary by up to ${\sim}30\%$ between seeds,
and only $0\text{--}15\%$ of a group's latents have a partner at decoder
$|\cos|>0.9$ in another seed's group for the same observable. In an
exploratory comparison (not energy matched), the replicated groups smaller than
the 128-dimensional residual stream nevertheless span subspaces that overlap
more than random groups of the same sizes (mean squared cosine of the principal
angles above all or nearly all of 999 random pairs, $p\le 0.002$), with one
exception: jet mass at $L=1024,k=8$, whose $114\text{--}119$ latents nearly
span all 128 dimensions ($p=0.015\text{--}0.834$). Replication
is therefore carried by partly shared subspaces, not by matching latents. All
seeds share the model, jets and activations, so this tests sensitivity to the
SAE initialisation only.

Together, LEACE (\hook{cls\_blocks.1}) and group ablation (\hook{blocks.7})
give a consistent picture. First, the two probes complement each
other: of the four observables whose linear erasure at \hook{cls\_blocks.1} was
incomplete ($m_J$, $\mSD$, $w$, $\pTD$), group ablation at \hook{blocks.7}
supports jet mass in every configuration and seed, with the largest losses;
$\mSD$ and $w$ replicate only at $L=1024,k=8$, and the seed-0 $\pTD$ claims do
not replicate. Second, the
two probes differ on multiplicity: erasing $\nconst$ at \hook{cls\_blocks.1} is
claimed, whereas multiplicity-aligned groups at \hook{blocks.7} cost less than
their nulls ($p\ge 0.993$), mostly with twice the nulls' decoded energy. No rule
tests losses below the null, so this is an observation; one untested reading is
that particle counting is formed during class-attention pooling.

%==============================================================================
\section{Infrared and collinear response}
\label{sec:irc}
%==============================================================================

Sections~\ref{sec:sae_results} and~\ref{sec:concept} identify the observables
\partkin{} uses by comparing its representation with observables we chose. IRC
safety can instead be tested directly: an IRC-safe quantity is unchanged when a
particle is split into two exactly collinear particles, and when a particle of
vanishing energy is added. We pre-registered such a test before computing any
response (Appendix~\ref{app:protocol}) and applied it to the $50\mathrm{k}$
evaluation jets of Section~\ref{sec:concept} (one jet already at the
128-particle limit is excluded).

\paragraph{Perturbations.} \emph{Collinear:} the hardest constituent is split
with momentum fraction $z=0.5$, exactly collinearly (the test point) and at
opening angles $\delta=10^{-3}, 10^{-2}, 10^{-1}$. Because JetClass
four-vectors carry generator masses that \partkin{} uses
(Section~\ref{sec:practical}), each daughter keeps the parent's rest mass; an
exact four-momentum split (daughter masses $z\,m$) gives the same
classifications. \emph{Soft:} one massless particle with a fraction
$f\in\{10^{-5}, 10^{-4}, 10^{-3}, 10^{-2}\}$ of the jet energy, uniform within
$\Delta R<0.8$ of the jet axis. A random permutation of the constituents, to
which \texttt{ParT} is invariant, sets each quantity's numerical noise floor.

\paragraph{Quantities and classification.} For each quantity $y$ we measure the
response $R_y=\langle|\Delta y|\rangle/\sigma_y$, with $\sigma_y$ its spread
over the unperturbed jets: the ten logits; linear readouts (ridge probes fitted
on the $200\mathrm{k}$ fit jets, held-out $R^2=0.55\text{--}0.93$) of $m_J$,
$\mSD$, $w$, $\tau_{21}$, $\tau_{32}$, $\nconst$, $\pTD$ and the 101 log EFPs
from the jet vector at \hook{cls\_blocks.1}; the per-jet summed activation of
every eligible seed-0 latent of the three gated \hook{blocks.7} SAEs and of
both controls; and the observables themselves. The scale is set by
multiplicity: adding one particle moves $\nconst$ by $R_{\mathrm{ref}}=0.066$.
A quantity is \emph{unsafe-like} if its response at the test point exceeds
$0.1\,R_{\mathrm{ref}}$ at $95\%$ bootstrap confidence~\cite{Efron:1979boot} (for soft emission also
requiring that it does not fall with $f$), and \emph{safe-like} if it stays
below $\max(10\times\text{floor}, 0.01\,R_{\mathrm{ref}})$ or (for soft
emission) falls at least as $f^{0.5}$. As a pre-registered check, these rules
classify the observables correctly: the EFPs, $m_J$ and $\mSD$ are safe-like
under both perturbations, $\nconst$ is unsafe-like under both, and $\pTD$ is
collinear-unsafe but infrared-safe.

\begin{figure*}[tbp]
\centering
\includegraphics[width=\textwidth]{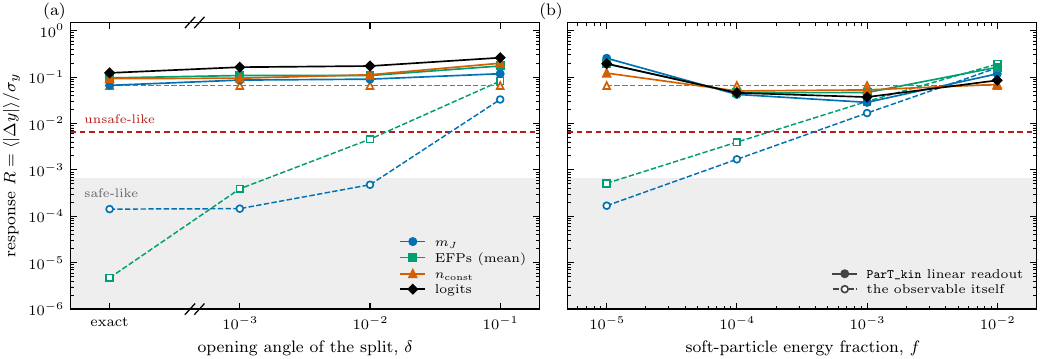}
\caption{\textbf{Infrared and collinear response of \partkin{} on $50\mathrm{k}$
held-out jets.} The response $R=\langle|\Delta y|\rangle/\sigma_y$ of the logits
(black), of linear readouts of jet observables from the jet vector at
\hook{cls\_blocks.1} (filled, solid), and of the observables themselves (open,
dashed) under (a)~splitting the hardest constituent at opening angle $\delta$
(``exact'': $\delta=0$, the test point) and (b)~adding a massless particle with
energy fraction $f$. Colours distinguish $m_J$, the EFPs (mean over the 101 log
EFPs) and $\nconst$; the axis break marks that the exact split ($\delta=0$) is
not on the logarithmic scale. Below the dashed red line ($0.1\times$ the
response of $\nconst$ to one extra particle) nothing is classified unsafe; the
grey region is safe-like. The IRC-safe
observables fall towards zero in both limits, while their readouts stay at the
level of the multiplicity readout. All readouts are compared with their
observables in Table~\ref{tab:irc}.}
\label{fig:irc}
\end{figure*}

\begin{table*}[t]
\centering
\caption{Response $R=\langle|\Delta y|\rangle/\sigma_y$ of each observable and
of its linear readout from the jet vector at \hook{cls\_blocks.1} at the two
test points ($49{,}999$ held-out jets): an exact collinear split of the hardest
constituent (daughters keep the parent's rest mass) and one massless particle
with $f=10^{-5}$ of the jet energy. \textbf{Bold}: unsafe-like (lower $95\%$
bound above $0.1\,R_{\mathrm{ref}}=0.0066$); all other entries are safe-like.
EFPs: mean over the 101 log EFPs.}
\label{tab:irc}
\begin{tabular}{llcccc}
\toprule
& & \multicolumn{2}{c}{Exact collinear split} & \multicolumn{2}{c}{Soft particle, $f=10^{-5}$} \\
\cmidrule(lr){3-4}\cmidrule(lr){5-6}
Quantity & IRC class & Observable & Readout & Observable & Readout \\
\midrule
$m_J$       & safe         & $1.4\times10^{-4}$ & \textbf{0.066} & $1.7\times10^{-4}$ & \textbf{0.255} \\
$\mSD$      & safe         & $2.2\times10^{-4}$ & \textbf{0.073} & $3.0\times10^{-5}$ & \textbf{0.170} \\
$w$         & safe         & $6.7\times10^{-6}$ & \textbf{0.090} & $5.8\times10^{-5}$ & \textbf{0.167} \\
EFPs        & safe         & $4.7\times10^{-6}$ & \textbf{0.096} & $5.1\times10^{-4}$ & \textbf{0.192} \\
$\tau_{21}$ & Sudakov safe & $1.4\times10^{-5}$ & \textbf{0.116} & $1.8\times10^{-4}$ & \textbf{0.194} \\
$\tau_{32}$ & Sudakov safe & $8.4\times10^{-5}$ & \textbf{0.159} & $6.0\times10^{-4}$ & \textbf{0.276} \\
$\pTD$      & unsafe       & \textbf{0.421}     & \textbf{0.276} & $2.6\times10^{-5}$ & \textbf{0.156} \\
$\nconst$   & unsafe       & \textbf{0.066}     & \textbf{0.093} & \textbf{0.066}     & \textbf{0.122} \\
\midrule
Logits      &              & --                 & \textbf{0.123} & --                 & \textbf{0.194} \\
\bottomrule
\end{tabular}
\end{table*}

\paragraph{Results.} Figure~\ref{fig:irc} and Table~\ref{tab:irc} summarise the
outcome. The logits are unsafe-like under both perturbations, as expected for
an architecture in which every particle is a token. An exact collinear split
changes the prediction for $7.3\%$ of jets (accuracy $-0.022$), and one massless
particle at $f=10^{-5}$ does so for $11.3\%$ ($-0.025$).

The readouts of IRC-safe observables are not IRC safe. Every linear readout is unsafe-like under both perturbations and in both mass
    treatments (Table~\ref{tab:irc}). The multiplicity readout behaves like
    multiplicity, but the readouts of IRC-safe observables respond just as
    strongly: under an exact collinear split the jet-mass readout moves by
    $R=0.066$ while $m_J$ itself moves by $1.4\times10^{-4}$. Only the readouts of $\nconst$ (both perturbations) and of $\pTD$ (collinear
split) match their observables.

Neither limit is smooth. The logit response is larger at
    $\delta=10^{-3}$ ($0.163$) than at the exact split ($0.123$), and the soft
    response \emph{grows} as $f\to 0$. At $f=10^{-5}$ the added particle
    carries $\mathcal{O}(10)$~MeV, at the clipping edge of the logarithmic
    $p_T$ input, so this limit probes extrapolation outside the training
distribution rather than the treatment of physical soft radiation. A large
response to one soft neutral particle was found independently in a jet
foundation model and traced to pretraining on datasets with different
reconstruction thresholds~\cite{Bonnet-Guerrini:2026mwq}. At
    $f=10^{-4}\text{--}10^{-3}$ the readouts still move by $R\approx
    0.02\text{--}0.05$, three to eight times the unsafe threshold.

For the SAE latents the test does not discriminate. Under the primary perturbations,
    between $97\%$ and $100\%$ of eligible latents are unsafe-like in every
    dictionary, trained and control alike, and none is safe-like. The one formally rejected population
    comparison ($L=4096,k=16$, collinear) runs in opposite directions against
    the two controls ($99.7\%$ against $98.3\%$ and $100\%$), and the
    alignment test has no power, so neither supports a claim. This is partly
    by construction: a latent's activation summed over particles responds to
    the number of particles as multiplicity does.

A per-particle test removes this limitation. An IRC-safe per-particle term,
such as one term of an EFP, is additive under an exact collinear split: the
two daughters' activations add up to the parent's. For every
\hook{blocks.7} latent that fires at the split particle in at least 200 jets
(159, 504 and 730 latents in the three dictionaries), we pre-registered an
additivity residual $A=\langle|z_{d_1}+z_{d_2}-z_p|\rangle/\langle z_p\rangle$,
a copying residual $C$ (each daughter keeps the parent's activation) and a
spectator change $S$ on the unchanged particles, calling a latent additive if
$A\le0.1$ and $S\le0.1$. No latent is additive, in any trained or control
dictionary; the smallest $A$ of a trained latent is $0.30$ (medians
$1.55\text{--}1.63$), so the outcome does not depend on the spectator condition.
A few random-init latents copy the parent (3, 8 and 28), none in the trained or
shuffled dictionaries. In the trained SAEs the split changes \emph{which}
latents fire rather than dividing their activation: a latent fires on a
daughter but not on the parent in many jets. Neither pre-registered rule
(trained versus controls; IRC-safe versus other aligned latents) can be claimed,
since no latent is additive.

\paragraph{Reading.} \partkin{} relies on IRC-safe information
(Section~\ref{sec:concept}), yet under an exact collinear split or one soft
particle its internal linear representation of jet mass, of the EFPs and of the
$N$-subjettiness ratios moves as much as its representation of multiplicity.
The tagger's notion of ``mass'' is thus entangled with particle counting, which hadronisation and detector
simulation control; IRC-safe architectures such as Energy Flow
Networks~\cite{Komiske:2018cqr} and energy-weighted message
passing~\cite{Konar:2021zdg} avoid this by construction. Because the tests
act on Delphes-level constituents, ``safe-like'' here means a vanishing
response over the tested range, not IRC safety to all orders.

%==============================================================================
\section{Same tests on \texorpdfstring{\partfull{}}{ParT\_full}}
\label{sec:full}
%==============================================================================

Are these results specific to the kinematics-only checkpoint? \partfull{} has
the same architecture and is trained on the same jets with, in addition,
particle charge, five PID flags and track impact parameters. We repeated three
tests on it unchanged, with rules fixed before any of its numbers existed:
LEACE at \hook{cls\_blocks.1}, the IRC response of the logits and readouts (no
SAEs were trained on \partfull{}), and the removal of generator masses. In the
IRC test the collinear daughters inherit the parent's charge, PID flags and
track parameters, and the soft particle is a photon. Our input pipeline reproduces its validated accuracy ($0.8595$ on the
$30\mathrm{k}$ validation jets; Section~\ref{sec:practical}). Table~\ref{tab:full} compares the two checkpoints.

\begin{table}[t]
\centering
\caption{\partkin{} versus \partfull{} under the same pre-registered tests
($50\mathrm{k}$ evaluation jets). LEACE: accuracy loss at \hook{cls\_blocks.1}
(\textbf{bold}: claimed; others pass BH but fail the completeness gate; null
means $0.03\text{--}0.05$). IRC: response $R$ of the logits and range over the
eight readouts; ``flips'': jets whose predicted class changes.}
\label{tab:full}
\begin{tabular}{lcc}
\toprule
 & \partkin{} & \partfull{} \\
\midrule
Accuracy                          & 0.746 & 0.860 \\
\midrule
\multicolumn{3}{l}{\emph{LEACE loss at \hook{cls\_blocks.1}}} \\
All known observables             & \textbf{0.333} & \textbf{0.301} \\
EFPs                              & \textbf{0.252} & \textbf{0.206} \\
$m_J$ / $\mSD$                    & 0.268 / 0.243 & 0.226 / 0.228 \\
$w$                               & 0.213 & 0.214 \\
$\tau_{21}$ / $\tau_{32}$         & \textbf{0.170} / \textbf{0.123} & \textbf{0.168} / \textbf{0.113} \\
$\nconst$ / $\pTD$                & \textbf{0.144} / 0.123 & \textbf{0.109} / 0.086 \\
Left after erasing all known      & 0.413 & 0.559 \\
\midrule
\multicolumn{3}{l}{\emph{IRC: exact collinear split}} \\
Logits $R$ (flips)                & 0.123 (7.3\%) & 0.065 (2.6\%) \\
Readouts $R$                      & $0.066\text{--}0.276$ & $0.083\text{--}0.161$ \\
\multicolumn{3}{l}{\emph{IRC: soft particle, $f=10^{-5}$}} \\
Logits $R$ (flips)                & 0.194 (11.3\%) & 0.012 (0.4\%) \\
Readouts $R$                      & $0.122\text{--}0.276$ & $0.016\text{--}0.031$ \\
\midrule
\multicolumn{3}{l}{\emph{Generator masses ($\Delta$Acc)}} \\
All massless                      & $-0.106$ & $-0.018$ \\
Permuted (control)                & $-0.115$ & $-0.026$ \\
\bottomrule
\end{tabular}
\end{table}

\paragraph{Concept erasure.} Calibration passes (jet $\eta$: $p=0.874$; jet
$\phi$: $p=0.157$), and the same five concepts are claimed as for \partkin{}:
the EFPs, $\tau_{21}$, $\tau_{32}$, multiplicity and all known observables
together, with $m_J$, $\mSD$, $w$ and $\pTD$ again failing the completeness
gate. The losses are similar. Erasing all known observables leaves $0.559$
(\partkin{}: $0.413$), consistent with \partfull{} drawing more on information
outside these observables, such as the impact parameters and PID flags.

\paragraph{IRC response.} The pre-registered reference check passes, and the
logits and all eight readouts are unsafe-like under the exact collinear split
and the soft photon, as for \partkin{}; again only the multiplicity readout
(both perturbations) and the $\pTD$ readout (collinear) match their observables.
Under the collinear split the readouts move about as much as in \partkin{}.
The soft response is about an order of magnitude smaller and falls slowly as $f\to0$, whereas
that of \partkin{} grows, so the large soft response of \partkin{} at
$f=10^{-5}$ is mostly specific to that checkpoint. With a charged-pion soft
particle (secondary mode) the logits are undetermined rather than unsafe-like.

\paragraph{Generator masses.} Removing the masses costs \partfull{} $0.018$ in
accuracy, a sixth of the cost for \partkin{}, with claimed recall losses in six
classes (largest $H\to gg$, $-0.060$); the permuted control costs $0.026$. Both
registered expectations held: removing the masses costs \partfull{} less,
consistent with explicit PID taking over from the mass cue, and the IRC classes
of the logits and readouts are the same as for \partkin{} in the primary
perturbations. The finding that IRC-safe information is stored in an
IRC-unsafe way is therefore not a consequence of the missing PID inputs.

%==============================================================================
\section{Discussion: physical synthesis and limitations}
\label{sec:discussion}
%==============================================================================

Table~\ref{tab:synthesis} consolidates the findings of all analyses,
grouped by physical regime.

\begin{table*}[t]
\centering
\caption{Synthesis of the analyses of \partkin{} on JetClass (last row:
\partfull{}), separating
claims under the pre-registered rules from diagnostic observations and withheld
tests. The population-level SAE result is pooled over all latents and is not
broken down by observable.}
\label{tab:synthesis}
\fitwidth{%
\begin{tabular}{llllll}
\toprule
Physics category & Per-latent SAE (pop.) & Single-latent FDR & LEACE (\hook{cls\_blocks.1}) & SAE group abl.\ (\hook{blocks.7}) & IRC response (Sec.~\ref{sec:irc}) \\
\midrule
IRC-safe mass \& shape & pooled excess:   & 0 certified & Large drop ($0.21\text{--}0.27$), & \textbf{$m_J$: 3/3 configs, all seeds}; & readouts unsafe-like, \\
($m_J, \mSD, w$)       & (a) $\times20$, (b) $\times3.6$ &  & incomplete (withheld)            & $w, \mSD$ replicated (1/3) & observables safe-like \\
\midrule
Prongness \& EFPs      & (vs.\ shuffled,  & 0 certified & \textbf{EFPs ($0.252$), $\tau_{21}$ ($0.170$),} & \textbf{$\tau_2$ replicated (2/3)}; & readouts unsafe-like, \\
($\tau_N$, EFPs; $\tau_{21},\tau_{32}$ Sudakov safe) & \hook{blocks.7}) & & \textbf{$\tau_{32}$ ($0.123$) claimed} & $\tau_3$ replicated (1/3); $\tau_{21}$ not & observables safe-like \\
\midrule
IRC-unsafe fragm.      &                 & 0 certified & \textbf{$\nconst$ ($0.144$) claimed}; & $\pTD$: seed 0 only, & $\nconst$ readout matches; \\
($\nconst$, $\pTD$)    &                 &             & $\pTD$ ($0.123$) incomplete      & not replicated; $\nconst$ never & $\pTD$ (collinear only) \\
\midrule
Hidden particle-mass   & 94/112 \hook{blocks.7} & 0 certified & Not directly erased & $n<10$ or not rejected & -- \\
PID (muons)            & muon-specific   & (best case: 1) & (lepton classes drop $0.50\text{--}0.58$) & ($p\ge 0.108$) & \\
 & \multicolumn{5}{l}{\emph{Input test (Sec.~\ref{sec:muons}):} masses removed $-0.106$; muons only $-0.016$; $K/p\to\pi$ $-0.022$ (largest loss $H\to b\bar b$)} \\
\midrule
Symmetry controls      & Shuffled \& rand.-init & 0 false disc. & \textbf{Pass} ($\eta: 0.945$, $\phi: 0.122$); & \textbf{Pass all 3 configs} & \textbf{Reference check} \\
($\eta$, $\phi$, nulls)& calibrated ($\le 1.1\%$) & (BH/BY)     & \hook{blocks.7} $\phi$ fails (no claims) & ($\eta, \phi$: $p\ge 0.257$) & \textbf{passes}; SAE test no power; \\
& & & & & per particle: 0 additive latents \\
\midrule
\partfull{} (Sec.~\ref{sec:full}) & \multicolumn{5}{l}{Same five LEACE claims; same IRC classes (soft response $\sim\!10\times$ smaller); masses removed: $-0.018$ (\partkin{}: $-0.106$)} \\
\bottomrule
\end{tabular}}
\end{table*}

\paragraph{Population evidence without identifiable atoms.}
For mechanistic interpretability in particle physics, the central lesson is the
contrast between population-level and single-latent evidence: the trained model
organises its particle representations around collider physics far beyond both
controls, yet no individual SAE atom is identifiable across seeds, at any width,
sparsity or training-set size tested (Section~\ref{sec:sae_results}).
\emph{Subspaces and feature groups}, rather than individual SAE atoms, are the
units for which causal statements could be made here.

\paragraph{Perturbative versus non-perturbative physics reliance.}
Table~\ref{tab:synthesis} shows that \partkin{} draws on both sides of the QCD
divide: IRC-safe and Sudakov-safe substructure (EFPs, $\tau_{21}$, $\tau_{32}$,
jet mass, $\tau_2$) carries much of the classification, and IRC-unsafe
fragmentation (multiplicity, at the jet vector) is used as well. Generator-level particle masses
are both represented (muon-specific features) and used: removing them costs
$0.106$ in accuracy, the muon and kaon/proton masses alone $0.016$ and $0.022$. Reliance on IRC-unsafe information suggests that such taggers may be sensitive
to parton-shower, hadronisation and particle-mass modelling, as IRC-unsafe
quark/gluon observables are~\cite{Gras:2017jty}; a Pythia versus
Herwig~\cite{Bellm:2015jjp} comparison (not done here) would test this directly. The direct IRC
test of Section~\ref{sec:irc} sharpens this: even the IRC-safe information is
encoded in an IRC-unsafe way, so collinear splittings and soft particles move
the tagger's representation of jet mass and the EFPs as much as its
representation of multiplicity, and no SAE latent splits its activation
additively between collinear daughters. The same holds for \partfull{}
(Section~\ref{sec:full}), so it does not come from the missing PID inputs. This argues for auditing tagger
representations against IRC safety and generator controls, and for IRC-safe
architectures~\cite{Komiske:2018cqr,Konar:2021zdg} or targeted model
edits~\cite{Bonnet-Guerrini:2026mwq}.

\paragraph{Connection to measurements.}
Collider measurements point the same way. ATLAS finds that, among
machine-learning top taggers evaluated on ATLAS-simulated jets, the most
performant algorithms carry the largest systematic
uncertainties~\cite{ATLAS:2024rua}. Measurements of quark- and gluon-jet
substructure~\cite{CMS:2021iwu}, of the charged-particle multiplicity inside
jets~\cite{ATLAS:2016vxz}, and of the Lund jet plane~\cite{ATLAS:2020bbn} show
that generators still do not describe the full radiation pattern of jets, in
particular its non-perturbative part, whereas groomed observables such as the
soft-drop jet mass are well described by resummed calculations and parton
showers~\cite{ATLAS:2019mgf}. Our results suggest a mechanism for this trade-off
in one tagger: \partkin{} relies on IRC-unsafe multiplicity and encodes even its
IRC-safe information in a way that tracks particle counting, so its output
inherits sensitivity to the modelling of hadronisation and detector response.
Testing this link directly, by applying the analyses presented here to taggers
with estimated systematic uncertainties, such as those trained on the public
ATLAS datasets~\cite{ATLAS:2024rua}, is a natural next step.

\paragraph{What the controls caught.}
Without controls that can fail, each of the following would have produced a
false positive: FVU and replication were matched or exceeded by untrained
and shuffled networks~\cite{Heap:2025random}; the isotropic causal null passed
up to $63\%$ of ordinary latents;
token-level LEACE failed its azimuthal-symmetry check at \hook{blocks.7}; four
\hook{cls\_blocks.1} erasures left residual linear probes above the completeness
limit; and the $\tau_1$ SAE group at $L=4096,k=16$ owed its large drop in part
to carrying twice the decoded energy of its nulls. A practical lesson for
dictionary studies follows: because
permutation $p$-values cannot fall below $1/(N_{\mathrm{null}}+1)$~\cite{Phipson:2010perm}, Monte
Carlo resolution must be planned jointly with the $\sim 10^4$-latent BH family
size from the outset.

\paragraph{Limitations and next steps.}
This study investigates one architecture on one Delphes-simulated Pythia
dataset (JetClass), with $40\mathrm{k}\text{--}50\mathrm{k}$ held-out jets. SAEs
were trained only for \partkin{}; \partfull{} was tested with concept erasure,
the IRC response and mass removal. Frequency matching in large control pools is
approximate; latents within an SAE are statistically dependent (addressed via
BY checks and per-configuration consistency); post-hoc amendments were logged
and calibrated before viewing real outputs but remain post hoc
(Appendix~\ref{app:protocol}); LEACE tests linear extractability only; and
group ablation uses three SAE seeds per configuration at a single hook. The IRC
tests act on Delphes-level constituents, and additivity under an exact
collinear split is necessary, not sufficient, for a latent to be IRC safe. The
classifiers on the known observables had not saturated, so what \partkin{}
uses beyond these observables is bounded from above, not measured. Two physics
follow-ups stand out: quantifying latent and subspace drift across parton-shower
and hadronisation generators (Pythia versus Herwig), and characterising any
information the tagger uses beyond the known observables.

%==============================================================================
\section{Conclusions}
\label{sec:conclusions}
%==============================================================================

We asked what the public kinematic Particle Transformer (\partkin{}) knows
about QCD, using top-$k$ sparse autoencoders, linear concept erasure,
energy-matched group ablation, and direct infrared and collinear perturbations
on JetClass, each tested against matched null controls, together with
classifiers on the known observables and the removal of particle masses from
the inputs. Returning to the
three physics questions posed in Section~\ref{sec:intro}, our measurements
support five conclusions:
\begin{enumerate}
  \item Learned physics is evident in the population of SAE latents, not in
    single atoms. Causal and physics-aligned latents are far more common in the
    trained \partkin{} than in the untrained and shuffled controls, yet no
    individual latent survives multiple-testing correction on all three
    criteria, even with $4{,}999$ null draws or $10^6$ training jets, because
    latents do not replicate across SAE seeds.
  \item \partkin{} causally relies on IRC-safe and Sudakov-safe substructure and
    on IRC-unsafe fragmentation. At the jet vector, erasing the EFPs,
    $\tau_{21}$, $\tau_{32}$ or multiplicity costs far more accuracy than
    matched random erasures; at the last particle block, ablating the latents
    aligned with jet mass does so in every configuration and SAE seed, and
    those aligned with $\tau_2$ in two configurations.
  \item These encodings are not IRC safe. An exact collinear split or one soft
    particle moves the linear readouts of IRC-safe observables as much as the
    multiplicity readout ($R=0.07\text{--}0.28$), while the observables
    themselves move by at most $6\times10^{-4}$, and no SAE latent is additive
    under the split. \partfull{}, trained with PID inputs, behaves the same
    way, with a smaller soft response.
  \item What \partkin{} uses beyond the known observables is bounded, not
    measured: $0.413$ accuracy is left after linearly erasing them, and a
    neural network trained on them alone reaches $0.600$ against $0.746$, still
    improving with more jets.
  \item ``Kinematics-only'' inputs carry generator particle masses, from which
    \partkin{} builds muon-specific features; removing the masses costs $0.106$
    in accuracy (\partfull{}: $0.018$). The public checkpoints require \texttt{weaver-core}~0.4.x.
\end{enumerate}

\partkin{} therefore uses jet mass and $\tau_2$, Energy Flow Polynomials, the
Sudakov-safe ratios $\tau_{21}$ and $\tau_{32}$, and constituent multiplicity.
It stores this knowledge across directions of its representation rather than in
individual SAE features, and not in an IRC-safe way; whether it uses anything
beyond the known observables remains open. Reaching these statements required
symmetry controls, matched nulls, completeness checks and pre-registered rules,
which standard SAE diagnostics do not provide.

%==============================================================================
\section{Reproducibility and computational cost}
\label{sec:repro}
%==============================================================================

Analysis code, SLURM job scripts, the research plan, and the complete
timestamped decision log are available from the corresponding author upon
reasonable request. All training splits, model initialisations, and Monte Carlo
permutation seeds are fixed in code, and \texttt{weaver-core} is pinned to
version~0.4.17. Every GPU job ran on a single 45\,GB MIG slice of an NVIDIA~B200.
SLURM accounting records $35.6$~GPU-hours over the initial 97 jobs, including
tests and failed or cancelled runs. Of these, $9.4$~GPU-h went to the
$10^6$-jet follow-up, $7.2$~GPU-h to SAE group ablation, $6.4$~GPU-h to the
$4{,}999$-draw \hook{blocks.7} test, and $2.0$~GPU-h to LEACE concept erasure;
the remainder covers activation extraction, the SAE training sweeps, and the
initial criterion passes. Studies run afterwards added $0.24$~GPU-h (the IRC
test of Section~\ref{sec:irc}), $14.4$~GPU-h (group ablation for SAE seeds~1
and~2) and $0.85$~GPU-h (mass removal, the per-particle IRC test and the
\partfull{} tests; the classifiers on the known observables ran on CPU), for a
total of $51.1$~GPU-h.

\section*{Data availability}
All data used in this analysis are publicly available: the simulated JetClass
dataset~\cite{JetClass} and the pretrained Particle Transformer
weights~\cite{Qu:2022mxj}, described in Section~\ref{sec:setup}. No new data
were generated. The analysis code is available from the corresponding author
upon reasonable request.

\begin{acknowledgments}
We thank Shruti Naryal (School of Mathematical and Statistical Sciences, IIT
Mandi) for helpful discussions on transformers and sparse autoencoders. We acknowledge computational resources provided by the Indian Institute of
Technology Mandi. This work received no specific funding.
Part of the analysis code was written with the AI assistant Claude
(Anthropic); the authors reviewed all code, results and text and take full
responsibility for them.
\end{acknowledgments}

%==============================================================================
\appendix
\section{Pre-registration timeline, post-hoc amendments, and disclosures}
\label{app:protocol}
%==============================================================================

Every experimental design choice, calibration acceptance rule, and post-hoc
amendment was recorded in a timestamped decision log prior to examining the
corresponding real-model outcomes, with six explicit exceptions disclosed
below:
\begin{enumerate}
  \item \textbf{Per-particle SAE criterion (pre-registered before any
    per-particle SAE was trained):} Defined criteria~(a)--(c) with 19 isotropic
    causal null draws and frequency-matched 95th-percentile alignment
    thresholds. Both amendments below are explicitly post hoc, motivated by
    calibration failure or $p$-value floors rather than inspection of real
    latents, and the pre-registered counts (58 pairs) are reported alongside
    the amended results in Section~\ref{sec:perfeature}.
    \begin{itemize}
      \item \emph{Amendment 1 (causal calibration and FDR):} Replaced the
        miscalibrated isotropic null with $p_a = \max(p_{\mathrm{cov}},
        p_{\mathrm{shuf}})$, expressed (b) as a Bonferroni-corrected empirical
        $p$-value, and added BH/BY at $q=0.05$. Validated on control latents
        with a pre-set ceiling ($\le 7.5\%$ passing at $\alpha=0.05$) before
        opening real outputs.
      \item \emph{Amendment 2 (Monte Carlo resolution only):} Raised null draws
        and frequency-matched control pools at \hook{blocks.7} from $499$ to
        $4{,}999$ (control seeds 1--8) without altering the test statistic,
        nulls, $\alpha$, family, or $q$; re-calibrated on the full control
        families before opening real outputs.
    \end{itemize}
  \item \textbf{LEACE concept erasure (pre-registered before any full-scale
    run):} The probe-completeness gate (rule~2) and the between-jet covariance
    null ($W B W$) were added after a tiny CPU smoke test (400 fit / 200 eval
    jets per file, 19 nulls, numbers unused) revealed residual probes and
    failed negative controls under a uniform whitened null at token hooks.
    \emph{Disclosure~1:} The automated decision script applying the erasure
    rules was written after the \hook{cls\_blocks.1} numbers had appeared in
    the job log (and before any \hook{blocks.7} output existed). It applies BH
    over all nine physics concepts; the alternative reading (BH only over
    concepts passing completeness) yields the identical five claims at
    \hook{cls\_blocks.1} and zero claims at \hook{blocks.7}.
  \item \textbf{SAE group ablation (pre-registered before any group-ablation
    number existed):} \emph{Disclosure~2:} The design and rules were written
    after the LEACE results and the per-configuration SAE group sizes were
    known, without using any single-latent ablation column or group-ablation
    output. \emph{Disclosure~3:} The exploratory aligned-only null
    ($p_{\mathrm{al}}$) was added after reading the first GPU smoke run for
    timing and energy ratios (before the full run); it remains strictly outside
    all decision rules and BH families.
  \item \textbf{IRC response (pre-registered before any IRC response was
    computed):} Perturbations, quantities, thresholds, controls and decision
    rules were fixed in advance and not amended after the smoke test.
    \emph{Disclosure~4:} The design was written after the SAE,
    concept-erasure and seed-0 group-ablation results were known. The population quantity (per-jet summed latent
    activation) turned out to respond to the particle count by construction,
    which limits rules~3 and~4 (Section~\ref{sec:irc}); this is reported, not
    corrected post hoc.
  \item \textbf{Group ablation across SAE seeds (pre-registered before any
    seed-1 or seed-2 label or ablation number existed):} the group-ablation
    design and rules were reused unchanged, with one added cross-seed rule
    (claimed in at least two of three seeds); no amendment after the smoke
    test. \emph{Disclosure~5:} The study was designed after the seed-0 results
    were known, in order to test them.
  \item \textbf{Robustness studies (pre-registered together, before any of
    their numbers existed):} classifiers on the known observables, removal of
    generator masses, the per-particle IRC test of SAE latents, and the
    \partfull{} tests. Implementation choices left open by the plan, including
    the criterion for a flat learning curve, were logged before each full run;
    no amendment followed any smoke test. \emph{Disclosure~6:} These studies
    were designed after all earlier results were known, in order to test them.
\end{enumerate}

\FloatBarrier
\bibliographystyle{apsrev4-2}
\bibliography{refs}

\begin{thebibliography}{50}%
\makeatletter
\providecommand \@ifxundefined [1]{%
 \@ifx{#1\undefined}
}%
\providecommand \@ifnum [1]{%
 \ifnum #1\expandafter \@firstoftwo
 \else \expandafter \@secondoftwo
 \fi
}%
\providecommand \@ifx [1]{%
 \ifx #1\expandafter \@firstoftwo
 \else \expandafter \@secondoftwo
 \fi
}%
\providecommand \natexlab [1]{#1}%
\providecommand \enquote  [1]{``#1''}%
\providecommand \bibnamefont  [1]{#1}%
\providecommand \bibfnamefont [1]{#1}%
\providecommand \citenamefont [1]{#1}%
\providecommand \href@noop [0]{\@secondoftwo}%
\providecommand \href [0]{\begingroup \@sanitize@url \@href}%
\providecommand \@href[1]{\@@startlink{#1}\@@href}%
\providecommand \@@href[1]{\endgroup#1\@@endlink}%
\providecommand \@sanitize@url [0]{\catcode `\\12\catcode `\$12\catcode
  `\&12\catcode `\#12\catcode `\^12\catcode `\_12\catcode `\%12\relax}%
\providecommand \@@startlink[1]{}%
\providecommand \@@endlink[0]{}%
\providecommand \url  [0]{\begingroup\@sanitize@url \@url }%
\providecommand \@url [1]{\endgroup\@href {#1}{\urlprefix }}%
\providecommand \urlprefix  [0]{URL }%
\providecommand \Eprint [0]{\href }%
\providecommand \doibase [0]{https://doi.org/}%
\providecommand \selectlanguage [0]{\@gobble}%
\providecommand \bibinfo  [0]{\@secondoftwo}%
\providecommand \bibfield  [0]{\@secondoftwo}%
\providecommand \translation [1]{[#1]}%
\providecommand \BibitemOpen [0]{}%
\providecommand \bibitemStop [0]{}%
\providecommand \bibitemNoStop [0]{.\EOS\space}%
\providecommand \EOS [0]{\spacefactor3000\relax}%
\providecommand \BibitemShut  [1]{\csname bibitem#1\endcsname}%
\let\auto@bib@innerbib\@empty
%</preamble>
\bibitem [{\citenamefont {Larkoski}\ \emph {et~al.}(2020)\citenamefont
  {Larkoski}, \citenamefont {Moult},\ and\ \citenamefont
  {Nachman}}]{Larkoski:2017jix}%
  \BibitemOpen
  \bibfield  {author} {\bibinfo {author} {\bibfnamefont {A.~J.}\ \bibnamefont
  {Larkoski}}, \bibinfo {author} {\bibfnamefont {I.}~\bibnamefont {Moult}},\
  and\ \bibinfo {author} {\bibfnamefont {B.}~\bibnamefont {Nachman}},\
  }\href@noop {} {\bibfield  {journal} {\bibinfo  {journal} {Phys. Rept.}\
  }\textbf {\bibinfo {volume} {841}},\ \bibinfo {pages} {1} (\bibinfo {year}
  {2020})},\ \Eprint {https://arxiv.org/abs/1709.04464} {arXiv:1709.04464}
  \BibitemShut {NoStop}%
\bibitem [{\citenamefont {Larkoski}\ \emph {et~al.}(2014)\citenamefont
  {Larkoski}, \citenamefont {Marzani}, \citenamefont {Soyez},\ and\
  \citenamefont {Thaler}}]{Larkoski:2014wba}%
  \BibitemOpen
  \bibfield  {author} {\bibinfo {author} {\bibfnamefont {A.~J.}\ \bibnamefont
  {Larkoski}}, \bibinfo {author} {\bibfnamefont {S.}~\bibnamefont {Marzani}},
  \bibinfo {author} {\bibfnamefont {G.}~\bibnamefont {Soyez}},\ and\ \bibinfo
  {author} {\bibfnamefont {J.}~\bibnamefont {Thaler}},\ }\href@noop {}
  {\bibfield  {journal} {\bibinfo  {journal} {J. High Energy Phys.}\ }\textbf
  {\bibinfo {volume} {05}}\bibfield  {number} {\bibinfo  {number} { (2014)},\
  \bibinfo {pages} {146}},\ }\Eprint {https://arxiv.org/abs/1402.2657}
  {arXiv:1402.2657} \BibitemShut {NoStop}%
\bibitem [{\citenamefont {Thaler}\ and\ \citenamefont
  {Van~Tilburg}(2011)}]{Thaler:2010tr}%
  \BibitemOpen
  \bibfield  {author} {\bibinfo {author} {\bibfnamefont {J.}~\bibnamefont
  {Thaler}}\ and\ \bibinfo {author} {\bibfnamefont {K.}~\bibnamefont
  {Van~Tilburg}},\ }\href@noop {} {\bibfield  {journal} {\bibinfo  {journal}
  {J. High Energy Phys.}\ }\textbf {\bibinfo {volume} {03}}\bibfield  {number}
  {\bibinfo  {number} { (2011)},\ \bibinfo {pages} {015}},\ }\Eprint
  {https://arxiv.org/abs/1011.2268} {arXiv:1011.2268} \BibitemShut {NoStop}%
\bibitem [{\citenamefont {Larkoski}\ and\ \citenamefont
  {Thaler}(2013)}]{Larkoski:2013paa}%
  \BibitemOpen
  \bibfield  {author} {\bibinfo {author} {\bibfnamefont {A.~J.}\ \bibnamefont
  {Larkoski}}\ and\ \bibinfo {author} {\bibfnamefont {J.}~\bibnamefont
  {Thaler}},\ }\href {https://doi.org/10.1007/JHEP09(2013)137} {\bibfield
  {journal} {\bibinfo  {journal} {J. High Energy Phys.}\ }\textbf {\bibinfo
  {volume} {09}}\bibfield  {number} {\bibinfo  {number} { (2013)},\ \bibinfo
  {pages} {137}},\ }\Eprint {https://arxiv.org/abs/1307.1699} {arXiv:1307.1699}
  \BibitemShut {NoStop}%
\bibitem [{\citenamefont {Larkoski}\ \emph {et~al.}(2015)\citenamefont
  {Larkoski}, \citenamefont {Marzani},\ and\ \citenamefont
  {Thaler}}]{Larkoski:2015lea}%
  \BibitemOpen
  \bibfield  {author} {\bibinfo {author} {\bibfnamefont {A.~J.}\ \bibnamefont
  {Larkoski}}, \bibinfo {author} {\bibfnamefont {S.}~\bibnamefont {Marzani}},\
  and\ \bibinfo {author} {\bibfnamefont {J.}~\bibnamefont {Thaler}},\ }\href
  {https://doi.org/10.1103/PhysRevD.91.111501} {\bibfield  {journal} {\bibinfo
  {journal} {Phys. Rev. D}\ }\textbf {\bibinfo {volume} {91}},\ \bibinfo
  {pages} {111501} (\bibinfo {year} {2015})},\ \Eprint
  {https://arxiv.org/abs/1502.01719} {arXiv:1502.01719 [hep-ph]} \BibitemShut
  {NoStop}%
\bibitem [{\citenamefont {Komiske}\ \emph {et~al.}(2018)\citenamefont
  {Komiske}, \citenamefont {Metodiev},\ and\ \citenamefont
  {Thaler}}]{Komiske:2017aww}%
  \BibitemOpen
  \bibfield  {author} {\bibinfo {author} {\bibfnamefont {P.~T.}\ \bibnamefont
  {Komiske}}, \bibinfo {author} {\bibfnamefont {E.~M.}\ \bibnamefont
  {Metodiev}},\ and\ \bibinfo {author} {\bibfnamefont {J.}~\bibnamefont
  {Thaler}},\ }\href@noop {} {\bibfield  {journal} {\bibinfo  {journal} {J.
  High Energy Phys.}\ }\textbf {\bibinfo {volume} {04}}\bibfield  {number}
  {\bibinfo  {number} { (2018)},\ \bibinfo {pages} {013}},\ }\Eprint
  {https://arxiv.org/abs/1712.07124} {arXiv:1712.07124} \BibitemShut {NoStop}%
\bibitem [{\citenamefont {Gallicchio}\ and\ \citenamefont
  {Schwartz}(2011)}]{Gallicchio:2011xq}%
  \BibitemOpen
  \bibfield  {author} {\bibinfo {author} {\bibfnamefont {J.}~\bibnamefont
  {Gallicchio}}\ and\ \bibinfo {author} {\bibfnamefont {M.~D.}\ \bibnamefont
  {Schwartz}},\ }\href@noop {} {\bibfield  {journal} {\bibinfo  {journal}
  {Phys. Rev. Lett.}\ }\textbf {\bibinfo {volume} {107}},\ \bibinfo {pages}
  {172001} (\bibinfo {year} {2011})},\ \Eprint
  {https://arxiv.org/abs/1106.3076} {arXiv:1106.3076} \BibitemShut {NoStop}%
\bibitem [{\citenamefont {{CMS Collaboration}}(2013)}]{CMS:2013kfa}%
  \BibitemOpen
  \bibfield  {author} {\bibinfo {author} {\bibnamefont {{CMS Collaboration}}},\
  }\href@noop {} {\emph {\bibinfo {title} {{Performance of quark/gluon
  discrimination in 8 TeV $pp$ data}}}},\ \bibinfo {type} {Tech. Rep.}\
  \bibinfo {number} {CMS-PAS-JME-13-002}\ (\bibinfo  {institution} {CERN},\
  \bibinfo {year} {2013})\ \bibinfo {note}
  {\url{https://cds.cern.ch/record/1599732}}\BibitemShut {NoStop}%
\bibitem [{\citenamefont {Qu}\ \emph {et~al.}(2022{\natexlab{a}})\citenamefont
  {Qu}, \citenamefont {Li},\ and\ \citenamefont {Qian}}]{Qu:2022mxj}%
  \BibitemOpen
  \bibfield  {author} {\bibinfo {author} {\bibfnamefont {H.}~\bibnamefont
  {Qu}}, \bibinfo {author} {\bibfnamefont {C.}~\bibnamefont {Li}},\ and\
  \bibinfo {author} {\bibfnamefont {S.}~\bibnamefont {Qian}},\ }\href@noop {}
  {\bibinfo {title} {{Particle Transformer for Jet Tagging}}} (\bibinfo {year}
  {2022}{\natexlab{a}}),\ \Eprint {https://arxiv.org/abs/2202.03772}
  {arXiv:2202.03772} \BibitemShut {NoStop}%
\bibitem [{\citenamefont {Sirunyan}\ \emph {et~al.}(2020)\citenamefont
  {Sirunyan} \emph {et~al.}}]{CMS:2020poo}%
  \BibitemOpen
  \bibfield  {author} {\bibinfo {author} {\bibfnamefont {A.~M.}\ \bibnamefont
  {Sirunyan}} \emph {et~al.} (\bibinfo {collaboration} {CMS}),\ }\href
  {https://doi.org/10.1088/1748-0221/15/06/P06005} {\bibfield  {journal}
  {\bibinfo  {journal} {JINST}\ }\textbf {\bibinfo {volume} {15}}\bibfield
  {number} {\bibinfo  {number} { (2020)},\ \bibinfo {pages} {P06005}},\
  }\Eprint {https://arxiv.org/abs/2004.08262} {arXiv:2004.08262 [hep-ex]}
  \BibitemShut {NoStop}%
\bibitem [{\citenamefont {Faucett}\ \emph {et~al.}(2021)\citenamefont
  {Faucett}, \citenamefont {Thaler},\ and\ \citenamefont
  {Whiteson}}]{Faucett:2020vbu}%
  \BibitemOpen
  \bibfield  {author} {\bibinfo {author} {\bibfnamefont {T.}~\bibnamefont
  {Faucett}}, \bibinfo {author} {\bibfnamefont {J.}~\bibnamefont {Thaler}},\
  and\ \bibinfo {author} {\bibfnamefont {D.}~\bibnamefont {Whiteson}},\
  }\href@noop {} {\bibfield  {journal} {\bibinfo  {journal} {Phys. Rev. D}\
  }\textbf {\bibinfo {volume} {103}},\ \bibinfo {pages} {036020} (\bibinfo
  {year} {2021})},\ \Eprint {https://arxiv.org/abs/2010.11998}
  {arXiv:2010.11998} \BibitemShut {NoStop}%
\bibitem [{\citenamefont {Agarwal}\ \emph {et~al.}(2021)\citenamefont
  {Agarwal}, \citenamefont {Hay}, \citenamefont {Iashvili}, \citenamefont
  {Mannix}, \citenamefont {McLean} \emph {et~al.}}]{Agarwal:2020fpt}%
  \BibitemOpen
  \bibfield  {author} {\bibinfo {author} {\bibfnamefont {G.}~\bibnamefont
  {Agarwal}}, \bibinfo {author} {\bibfnamefont {L.}~\bibnamefont {Hay}},
  \bibinfo {author} {\bibfnamefont {I.}~\bibnamefont {Iashvili}}, \bibinfo
  {author} {\bibfnamefont {B.}~\bibnamefont {Mannix}}, \bibinfo {author}
  {\bibfnamefont {C.}~\bibnamefont {McLean}}, \emph {et~al.},\ }\href@noop {}
  {\bibfield  {journal} {\bibinfo  {journal} {J. High Energy Phys.}\ }\textbf
  {\bibinfo {volume} {05}}\bibfield  {number} {\bibinfo  {number} { (2021)},\
  \bibinfo {pages} {208}},\ }\Eprint {https://arxiv.org/abs/2011.13466}
  {arXiv:2011.13466} \BibitemShut {NoStop}%
\bibitem [{\citenamefont {Wang}\ \emph {et~al.}(2024)\citenamefont {Wang},
  \citenamefont {Gandrakota}, \citenamefont {Ngadiuba}, \citenamefont {Sahu},
  \citenamefont {Bhatnagar} \emph {et~al.}}]{Wang:2024interp}%
  \BibitemOpen
  \bibfield  {author} {\bibinfo {author} {\bibfnamefont {A.}~\bibnamefont
  {Wang}}, \bibinfo {author} {\bibfnamefont {A.}~\bibnamefont {Gandrakota}},
  \bibinfo {author} {\bibfnamefont {J.}~\bibnamefont {Ngadiuba}}, \bibinfo
  {author} {\bibfnamefont {V.}~\bibnamefont {Sahu}}, \bibinfo {author}
  {\bibfnamefont {P.}~\bibnamefont {Bhatnagar}}, \emph {et~al.},\ }\href@noop
  {} {\bibinfo {title} {{Interpreting Transformers for Jet Tagging}}} (\bibinfo
  {year} {2024}),\ \Eprint {https://arxiv.org/abs/2412.03673}
  {arXiv:2412.03673} \BibitemShut {NoStop}%
\bibitem [{\citenamefont {Legge}\ \emph {et~al.}(2025)\citenamefont {Legge},
  \citenamefont {Wang}, \citenamefont {Ortiz}, \citenamefont {Limouzi},
  \citenamefont {Zhao} \emph {et~al.}}]{Legge:2025cnm}%
  \BibitemOpen
  \bibfield  {author} {\bibinfo {author} {\bibfnamefont {T.}~\bibnamefont
  {Legge}}, \bibinfo {author} {\bibfnamefont {A.}~\bibnamefont {Wang}},
  \bibinfo {author} {\bibfnamefont {J.}~\bibnamefont {Ortiz}}, \bibinfo
  {author} {\bibfnamefont {V.}~\bibnamefont {Limouzi}}, \bibinfo {author}
  {\bibfnamefont {Z.}~\bibnamefont {Zhao}}, \emph {et~al.},\ }\href@noop {}
  {\bibinfo {title} {{Why Is Attention Sparse In Particle Transformer?}}},\
  \bibinfo {howpublished} {Machine Learning and the Physical Sciences workshop,
  NeurIPS 2025} (\bibinfo {year} {2025}),\ \Eprint
  {https://arxiv.org/abs/2512.00210} {arXiv:2512.00210} \BibitemShut {NoStop}%
\bibitem [{\citenamefont {Rai}\ and\ \citenamefont
  {Ganguly}(2026)}]{Rai:2026uzm}%
  \BibitemOpen
  \bibfield  {author} {\bibinfo {author} {\bibfnamefont {S.}~\bibnamefont
  {Rai}}\ and\ \bibinfo {author} {\bibfnamefont {S.}~\bibnamefont {Ganguly}},\
  }\href@noop {} {\bibinfo {title} {{Dissecting Jet-Tagger Through Mechanistic
  Interpretability}}} (\bibinfo {year} {2026}),\ \Eprint
  {https://arxiv.org/abs/2605.09881} {arXiv:2605.09881} \BibitemShut {NoStop}%
\bibitem [{\citenamefont {Agarwal}\ \emph {et~al.}(2026)\citenamefont
  {Agarwal}, \citenamefont {Khare},\ and\ \citenamefont
  {Kumar}}]{Agarwal:2026uqw}%
  \BibitemOpen
  \bibfield  {author} {\bibinfo {author} {\bibfnamefont {J.}~\bibnamefont
  {Agarwal}}, \bibinfo {author} {\bibfnamefont {S.}~\bibnamefont {Khare}},\
  and\ \bibinfo {author} {\bibfnamefont {D.}~\bibnamefont {Kumar}},\
  }\href@noop {} {\bibinfo {title} {{What Do Lorentz-Equivariant Jet Taggers
  Learn?}}} (\bibinfo {year} {2026}),\ \Eprint
  {https://arxiv.org/abs/2606.21790} {arXiv:2606.21790} \BibitemShut {NoStop}%
\bibitem [{\citenamefont {Laatu}\ \emph {et~al.}(2026)\citenamefont {Laatu},
  \citenamefont {Bal}, \citenamefont {Maier}, \citenamefont {Klute},\ and\
  \citenamefont {Spannowsky}}]{Laatu:2026quw}%
  \BibitemOpen
  \bibfield  {author} {\bibinfo {author} {\bibfnamefont {L.}~\bibnamefont
  {Laatu}}, \bibinfo {author} {\bibfnamefont {A.}~\bibnamefont {Bal}}, \bibinfo
  {author} {\bibfnamefont {B.}~\bibnamefont {Maier}}, \bibinfo {author}
  {\bibfnamefont {M.}~\bibnamefont {Klute}},\ and\ \bibinfo {author}
  {\bibfnamefont {M.}~\bibnamefont {Spannowsky}},\ }\href@noop {} {\bibinfo
  {title} {{Higher Order Component Attribution via Learned Surrogates in High
  Energy Physics}}} (\bibinfo {year} {2026}),\ \Eprint
  {https://arxiv.org/abs/2611.08656} {arXiv:2611.08656} \BibitemShut {NoStop}%
\bibitem [{\citenamefont {Calder{\'o}n}\ \emph {et~al.}(2026)\citenamefont
  {Calder{\'o}n}, \citenamefont {Kato}, \citenamefont {Timiryasov},\ and\
  \citenamefont {Ruchayskiy}}]{Calderon:2026qqj}%
  \BibitemOpen
  \bibfield  {author} {\bibinfo {author} {\bibfnamefont {K.~A.~U.}\
  \bibnamefont {Calder{\'o}n}}, \bibinfo {author} {\bibfnamefont
  {M.}~\bibnamefont {Kato}}, \bibinfo {author} {\bibfnamefont {I.}~\bibnamefont
  {Timiryasov}},\ and\ \bibinfo {author} {\bibfnamefont {O.}~\bibnamefont
  {Ruchayskiy}},\ }\href@noop {} {\bibinfo {title} {{Who Moved My $Z$ Boson?
  Revealing Resonance Features Inside a Neural Network}}} (\bibinfo {year}
  {2026}),\ \Eprint {https://arxiv.org/abs/2611.26291} {arXiv:2611.26291}
  \BibitemShut {NoStop}%
\bibitem [{\citenamefont {Gambhir}\ \emph {et~al.}(2026)\citenamefont
  {Gambhir}, \citenamefont {Lucie-Smith},\ and\ \citenamefont
  {Thaler}}]{Gambhir:2026interp}%
  \BibitemOpen
  \bibfield  {author} {\bibinfo {author} {\bibfnamefont {R.}~\bibnamefont
  {Gambhir}}, \bibinfo {author} {\bibfnamefont {L.}~\bibnamefont
  {Lucie-Smith}},\ and\ \bibinfo {author} {\bibfnamefont {J.}~\bibnamefont
  {Thaler}},\ }\href@noop {} {\bibinfo {title} {{Interpreting
  ``Interpretability'' and Explaining ``Explainability'' in Machine Learning in
  Physics}}} (\bibinfo {year} {2026}),\ \Eprint
  {https://arxiv.org/abs/2606.26228} {arXiv:2606.26228} \BibitemShut {NoStop}%
\bibitem [{\citenamefont {Cunningham}\ \emph {et~al.}(2023)\citenamefont
  {Cunningham}, \citenamefont {Ewart}, \citenamefont {Riggs}, \citenamefont
  {Huben},\ and\ \citenamefont {Sharkey}}]{Cunningham:2023sae}%
  \BibitemOpen
  \bibfield  {author} {\bibinfo {author} {\bibfnamefont {H.}~\bibnamefont
  {Cunningham}}, \bibinfo {author} {\bibfnamefont {A.}~\bibnamefont {Ewart}},
  \bibinfo {author} {\bibfnamefont {L.}~\bibnamefont {Riggs}}, \bibinfo
  {author} {\bibfnamefont {R.}~\bibnamefont {Huben}},\ and\ \bibinfo {author}
  {\bibfnamefont {L.}~\bibnamefont {Sharkey}},\ }\href@noop {} {\bibinfo
  {title} {{Sparse Autoencoders Find Highly Interpretable Features in Language
  Models}}} (\bibinfo {year} {2023}),\ \Eprint
  {https://arxiv.org/abs/2309.08600} {arXiv:2309.08600} \BibitemShut {NoStop}%
\bibitem [{\citenamefont {Gao}\ \emph {et~al.}(2024)\citenamefont {Gao},
  \citenamefont {Dupr{\'e}~la Tour}, \citenamefont {Tillman}, \citenamefont
  {Goh}, \citenamefont {Troll}, \citenamefont {Radford} \emph
  {et~al.}}]{Gao:2024sae}%
  \BibitemOpen
  \bibfield  {author} {\bibinfo {author} {\bibfnamefont {L.}~\bibnamefont
  {Gao}}, \bibinfo {author} {\bibfnamefont {T.}~\bibnamefont {Dupr{\'e}~la
  Tour}}, \bibinfo {author} {\bibfnamefont {H.}~\bibnamefont {Tillman}},
  \bibinfo {author} {\bibfnamefont {G.}~\bibnamefont {Goh}}, \bibinfo {author}
  {\bibfnamefont {R.}~\bibnamefont {Troll}}, \bibinfo {author} {\bibfnamefont
  {A.}~\bibnamefont {Radford}}, \emph {et~al.},\ }\href@noop {} {\bibinfo
  {title} {{Scaling and evaluating sparse autoencoders}}} (\bibinfo {year}
  {2024}),\ \Eprint {https://arxiv.org/abs/2406.04093} {arXiv:2406.04093}
  \BibitemShut {NoStop}%
\bibitem [{\citenamefont {Bonnet-Guerrini}\ \emph
  {et~al.}(2026{\natexlab{a}})\citenamefont {Bonnet-Guerrini}, \citenamefont
  {Ioannou-Nikolaides}, \citenamefont {Timiryasov},\ and\ \citenamefont
  {Piuri}}]{BonnetGuerrini:2026sae}%
  \BibitemOpen
  \bibfield  {author} {\bibinfo {author} {\bibfnamefont {R.}~\bibnamefont
  {Bonnet-Guerrini}}, \bibinfo {author} {\bibfnamefont {J.}~\bibnamefont
  {Ioannou-Nikolaides}}, \bibinfo {author} {\bibfnamefont {I.}~\bibnamefont
  {Timiryasov}},\ and\ \bibinfo {author} {\bibfnamefont {V.}~\bibnamefont
  {Piuri}},\ }\href@noop {} {\bibinfo {title} {{Finding and using interpretable
  latents in a neutrino foundation model with sparse autoencoders}}} (\bibinfo
  {year} {2026}{\natexlab{a}}),\ \Eprint {https://arxiv.org/abs/2608.26090}
  {arXiv:2608.26090} \BibitemShut {NoStop}%
\bibitem [{\citenamefont {Qi}\ and\ \citenamefont {Earls}(2026)}]{Qi:2026xmb}%
  \BibitemOpen
  \bibfield  {author} {\bibinfo {author} {\bibfnamefont {Z.}~\bibnamefont
  {Qi}}\ and\ \bibinfo {author} {\bibfnamefont {C.}~\bibnamefont {Earls}},\
  }\href@noop {} {\bibinfo {title} {{Mechanistic Interpretability and Causal
  Feature Steering of Neural Quantum States via Sparse Autoencoders}}}
  (\bibinfo {year} {2026}),\ \Eprint {https://arxiv.org/abs/2607.01336}
  {arXiv:2607.01336} \BibitemShut {NoStop}%
\bibitem [{\citenamefont {Bonnet-Guerrini}\ \emph
  {et~al.}(2026{\natexlab{b}})\citenamefont {Bonnet-Guerrini}, \citenamefont
  {Ioannou-Nikolaides},\ and\ \citenamefont
  {Petersen}}]{Bonnet-Guerrini:2026mwq}%
  \BibitemOpen
  \bibfield  {author} {\bibinfo {author} {\bibfnamefont {R.}~\bibnamefont
  {Bonnet-Guerrini}}, \bibinfo {author} {\bibfnamefont {J.}~\bibnamefont
  {Ioannou-Nikolaides}},\ and\ \bibinfo {author} {\bibfnamefont
  {T.}~\bibnamefont {Petersen}},\ }\href@noop {} {\bibinfo {title}
  {{Weight-space programming and mechanistic interpretability for model editing
  in particle physics}}} (\bibinfo {year} {2026}{\natexlab{b}}),\ \Eprint
  {https://arxiv.org/abs/2611.27079} {arXiv:2611.27079 [hep-ph]} \BibitemShut
  {NoStop}%
\bibitem [{\citenamefont {Heap}\ \emph {et~al.}(2025)\citenamefont {Heap},
  \citenamefont {Lawson}, \citenamefont {Farnik},\ and\ \citenamefont
  {Aitchison}}]{Heap:2025random}%
  \BibitemOpen
  \bibfield  {author} {\bibinfo {author} {\bibfnamefont {T.}~\bibnamefont
  {Heap}}, \bibinfo {author} {\bibfnamefont {T.}~\bibnamefont {Lawson}},
  \bibinfo {author} {\bibfnamefont {L.}~\bibnamefont {Farnik}},\ and\ \bibinfo
  {author} {\bibfnamefont {L.}~\bibnamefont {Aitchison}},\ }\href@noop {}
  {\bibinfo {title} {{Automated Interpretability Metrics Do Not Distinguish
  Trained and Random Transformers}}} (\bibinfo {year} {2025}),\ \Eprint
  {https://arxiv.org/abs/2501.17727} {arXiv:2501.17727} \BibitemShut {NoStop}%
\bibitem [{\citenamefont {Belrose}\ \emph {et~al.}(2023)\citenamefont
  {Belrose}, \citenamefont {Schneider-Joseph}, \citenamefont {Ravfogel},
  \citenamefont {Cotterell}, \citenamefont {Raff},\ and\ \citenamefont
  {Biderman}}]{Belrose:2023leace}%
  \BibitemOpen
  \bibfield  {author} {\bibinfo {author} {\bibfnamefont {N.}~\bibnamefont
  {Belrose}}, \bibinfo {author} {\bibfnamefont {D.}~\bibnamefont
  {Schneider-Joseph}}, \bibinfo {author} {\bibfnamefont {S.}~\bibnamefont
  {Ravfogel}}, \bibinfo {author} {\bibfnamefont {R.}~\bibnamefont {Cotterell}},
  \bibinfo {author} {\bibfnamefont {E.}~\bibnamefont {Raff}},\ and\ \bibinfo
  {author} {\bibfnamefont {S.}~\bibnamefont {Biderman}},\ }\href@noop {}
  {\bibinfo {title} {{LEACE: Perfect linear concept erasure in closed form}}}
  (\bibinfo {year} {2023}),\ \Eprint {https://arxiv.org/abs/2306.03819}
  {arXiv:2306.03819} \BibitemShut {NoStop}%
\bibitem [{\citenamefont {Qu}\ \emph {et~al.}(2022{\natexlab{b}})\citenamefont
  {Qu}, \citenamefont {Li},\ and\ \citenamefont {Qian}}]{JetClass}%
  \BibitemOpen
  \bibfield  {author} {\bibinfo {author} {\bibfnamefont {H.}~\bibnamefont
  {Qu}}, \bibinfo {author} {\bibfnamefont {C.}~\bibnamefont {Li}},\ and\
  \bibinfo {author} {\bibfnamefont {S.}~\bibnamefont {Qian}},\ }\href
  {https://doi.org/10.5281/zenodo.6619768} {\bibinfo {title} {{JetClass: A
  Large-Scale Dataset for Deep Learning in Jet Physics}}} (\bibinfo {year}
  {2022}{\natexlab{b}})\BibitemShut {NoStop}%
\bibitem [{\citenamefont {Alwall}\ \emph {et~al.}(2014)\citenamefont {Alwall},
  \citenamefont {Frederix}, \citenamefont {Frixione}, \citenamefont {Hirschi},
  \citenamefont {Maltoni}, \citenamefont {Mattelaer}, \citenamefont {Shao},
  \citenamefont {Stelzer}, \citenamefont {Torrielli},\ and\ \citenamefont
  {Zaro}}]{Alwall:2014hca}%
  \BibitemOpen
  \bibfield  {author} {\bibinfo {author} {\bibfnamefont {J.}~\bibnamefont
  {Alwall}}, \bibinfo {author} {\bibfnamefont {R.}~\bibnamefont {Frederix}},
  \bibinfo {author} {\bibfnamefont {S.}~\bibnamefont {Frixione}}, \bibinfo
  {author} {\bibfnamefont {V.}~\bibnamefont {Hirschi}}, \bibinfo {author}
  {\bibfnamefont {F.}~\bibnamefont {Maltoni}}, \bibinfo {author} {\bibfnamefont
  {O.}~\bibnamefont {Mattelaer}}, \bibinfo {author} {\bibfnamefont {H.~S.}\
  \bibnamefont {Shao}}, \bibinfo {author} {\bibfnamefont {T.}~\bibnamefont
  {Stelzer}}, \bibinfo {author} {\bibfnamefont {P.}~\bibnamefont {Torrielli}},\
  and\ \bibinfo {author} {\bibfnamefont {M.}~\bibnamefont {Zaro}},\ }\href
  {https://doi.org/10.1007/JHEP07(2014)079} {\bibfield  {journal} {\bibinfo
  {journal} {J. High Energy Phys.}\ }\textbf {\bibinfo {volume} {07}}\bibfield
  {number} {\bibinfo  {number} { (2014)},\ \bibinfo {pages} {079}},\ }\Eprint
  {https://arxiv.org/abs/1405.0301} {arXiv:1405.0301 [hep-ph]} \BibitemShut
  {NoStop}%
\bibitem [{\citenamefont {Sj{\"o}strand}\ \emph {et~al.}(2015)\citenamefont
  {Sj{\"o}strand}, \citenamefont {Ask}, \citenamefont {Christiansen},
  \citenamefont {Corke}, \citenamefont {Desai}, \citenamefont {Ilten},
  \citenamefont {Mrenna}, \citenamefont {Prestel}, \citenamefont {Rasmussen},\
  and\ \citenamefont {Skands}}]{Sjostrand:2014zea}%
  \BibitemOpen
  \bibfield  {author} {\bibinfo {author} {\bibfnamefont {T.}~\bibnamefont
  {Sj{\"o}strand}}, \bibinfo {author} {\bibfnamefont {S.}~\bibnamefont {Ask}},
  \bibinfo {author} {\bibfnamefont {J.~R.}\ \bibnamefont {Christiansen}},
  \bibinfo {author} {\bibfnamefont {R.}~\bibnamefont {Corke}}, \bibinfo
  {author} {\bibfnamefont {N.}~\bibnamefont {Desai}}, \bibinfo {author}
  {\bibfnamefont {P.}~\bibnamefont {Ilten}}, \bibinfo {author} {\bibfnamefont
  {S.}~\bibnamefont {Mrenna}}, \bibinfo {author} {\bibfnamefont
  {S.}~\bibnamefont {Prestel}}, \bibinfo {author} {\bibfnamefont {C.~O.}\
  \bibnamefont {Rasmussen}},\ and\ \bibinfo {author} {\bibfnamefont {P.~Z.}\
  \bibnamefont {Skands}},\ }\href {https://doi.org/10.1016/j.cpc.2015.01.024}
  {\bibfield  {journal} {\bibinfo  {journal} {Comput. Phys. Commun.}\ }\textbf
  {\bibinfo {volume} {191}},\ \bibinfo {pages} {159} (\bibinfo {year}
  {2015})},\ \Eprint {https://arxiv.org/abs/1410.3012} {arXiv:1410.3012
  [hep-ph]} \BibitemShut {NoStop}%
\bibitem [{\citenamefont {de~Favereau}\ \emph {et~al.}(2014)\citenamefont
  {de~Favereau}, \citenamefont {Delaere}, \citenamefont {Demin}, \citenamefont
  {Giammanco}, \citenamefont {Lema{\^\i}tre}, \citenamefont {Mertens},\ and\
  \citenamefont {Selvaggi}}]{deFavereau:2013fsa}%
  \BibitemOpen
  \bibfield  {author} {\bibinfo {author} {\bibfnamefont {J.}~\bibnamefont
  {de~Favereau}}, \bibinfo {author} {\bibfnamefont {C.}~\bibnamefont
  {Delaere}}, \bibinfo {author} {\bibfnamefont {P.}~\bibnamefont {Demin}},
  \bibinfo {author} {\bibfnamefont {A.}~\bibnamefont {Giammanco}}, \bibinfo
  {author} {\bibfnamefont {V.}~\bibnamefont {Lema{\^\i}tre}}, \bibinfo {author}
  {\bibfnamefont {A.}~\bibnamefont {Mertens}},\ and\ \bibinfo {author}
  {\bibfnamefont {M.}~\bibnamefont {Selvaggi}} (\bibinfo {collaboration}
  {DELPHES 3}),\ }\href {https://doi.org/10.1007/JHEP02(2014)057} {\bibfield
  {journal} {\bibinfo  {journal} {J. High Energy Phys.}\ }\textbf {\bibinfo
  {volume} {02}}\bibfield  {number} {\bibinfo  {number} { (2014)},\ \bibinfo
  {pages} {057}},\ }\Eprint {https://arxiv.org/abs/1307.6346} {arXiv:1307.6346
  [hep-ex]} \BibitemShut {NoStop}%
\bibitem [{\citenamefont {Cacciari}\ \emph {et~al.}(2008)\citenamefont
  {Cacciari}, \citenamefont {Salam},\ and\ \citenamefont
  {Soyez}}]{Cacciari:2008gp}%
  \BibitemOpen
  \bibfield  {author} {\bibinfo {author} {\bibfnamefont {M.}~\bibnamefont
  {Cacciari}}, \bibinfo {author} {\bibfnamefont {G.~P.}\ \bibnamefont
  {Salam}},\ and\ \bibinfo {author} {\bibfnamefont {G.}~\bibnamefont {Soyez}},\
  }\href@noop {} {\bibfield  {journal} {\bibinfo  {journal} {J. High Energy
  Phys.}\ }\textbf {\bibinfo {volume} {04}}\bibfield  {number} {\bibinfo
  {number} { (2008)},\ \bibinfo {pages} {063}},\ }\Eprint
  {https://arxiv.org/abs/0802.1189} {arXiv:0802.1189} \BibitemShut {NoStop}%
\bibitem [{\citenamefont {Qu}()}]{weaver}%
  \BibitemOpen
  \bibfield  {author} {\bibinfo {author} {\bibfnamefont {H.}~\bibnamefont
  {Qu}},\ }\href@noop {} {\bibinfo {title} {{weaver-core}}},\ \bibinfo {note}
  {\url{https://github.com/hqucms/weaver-core}; version 0.4.17 used here,
  accessed November 2026}\BibitemShut {NoStop}%
\bibitem [{\citenamefont {Besag}\ and\ \citenamefont
  {Clifford}(1991)}]{Besag:1991mc}%
  \BibitemOpen
  \bibfield  {author} {\bibinfo {author} {\bibfnamefont {J.}~\bibnamefont
  {Besag}}\ and\ \bibinfo {author} {\bibfnamefont {P.}~\bibnamefont
  {Clifford}},\ }\href {https://doi.org/10.2307/2337256} {\bibfield  {journal}
  {\bibinfo  {journal} {Biometrika}\ }\textbf {\bibinfo {volume} {78}},\
  \bibinfo {pages} {301} (\bibinfo {year} {1991})}\BibitemShut {NoStop}%
\bibitem [{\citenamefont {Benjamini}\ and\ \citenamefont
  {Hochberg}(1995)}]{Benjamini:1995fdr}%
  \BibitemOpen
  \bibfield  {author} {\bibinfo {author} {\bibfnamefont {Y.}~\bibnamefont
  {Benjamini}}\ and\ \bibinfo {author} {\bibfnamefont {Y.}~\bibnamefont
  {Hochberg}},\ }\href {https://doi.org/10.1111/j.2517-6161.1995.tb02031.x}
  {\bibfield  {journal} {\bibinfo  {journal} {J. R. Stat. Soc. B}\ }\textbf
  {\bibinfo {volume} {57}},\ \bibinfo {pages} {289} (\bibinfo {year}
  {1995})}\BibitemShut {NoStop}%
\bibitem [{\citenamefont {Benjamini}\ and\ \citenamefont
  {Yekutieli}(2001)}]{Benjamini:2001by}%
  \BibitemOpen
  \bibfield  {author} {\bibinfo {author} {\bibfnamefont {Y.}~\bibnamefont
  {Benjamini}}\ and\ \bibinfo {author} {\bibfnamefont {D.}~\bibnamefont
  {Yekutieli}},\ }\href {https://doi.org/10.1214/aos/1013699998} {\bibfield
  {journal} {\bibinfo  {journal} {Ann. Statist.}\ }\textbf {\bibinfo {volume}
  {29}},\ \bibinfo {pages} {1165} (\bibinfo {year} {2001})}\BibitemShut
  {NoStop}%
\bibitem [{\citenamefont {Newcombe}(1998)}]{Newcombe:1998ci}%
  \BibitemOpen
  \bibfield  {author} {\bibinfo {author} {\bibfnamefont {R.~G.}\ \bibnamefont
  {Newcombe}},\ }\href
  {https://doi.org/10.1002/(SICI)1097-0258(19980430)17:8<873::AID-SIM779>3.0.CO;2-I}
  {\bibfield  {journal} {\bibinfo  {journal} {Stat. Med.}\ }\textbf {\bibinfo
  {volume} {17}},\ \bibinfo {pages} {873} (\bibinfo {year} {1998})}\BibitemShut
  {NoStop}%
\bibitem [{\citenamefont {Elhage}\ \emph {et~al.}(2022)\citenamefont {Elhage},
  \citenamefont {Hume}, \citenamefont {Olsson}, \citenamefont {Schiefer},
  \citenamefont {Henighan}, \citenamefont {Kravec}, \citenamefont
  {Hatfield-Dodds}, \citenamefont {Lasenby}, \citenamefont {Drain},
  \citenamefont {Chen}, \citenamefont {Grosse}, \citenamefont {McCandlish},
  \citenamefont {Kaplan}, \citenamefont {Amodei}, \citenamefont {Wattenberg},\
  and\ \citenamefont {Olah}}]{Elhage:2022toy}%
  \BibitemOpen
  \bibfield  {author} {\bibinfo {author} {\bibfnamefont {N.}~\bibnamefont
  {Elhage}}, \bibinfo {author} {\bibfnamefont {T.}~\bibnamefont {Hume}},
  \bibinfo {author} {\bibfnamefont {C.}~\bibnamefont {Olsson}}, \bibinfo
  {author} {\bibfnamefont {N.}~\bibnamefont {Schiefer}}, \bibinfo {author}
  {\bibfnamefont {T.}~\bibnamefont {Henighan}}, \bibinfo {author}
  {\bibfnamefont {S.}~\bibnamefont {Kravec}}, \bibinfo {author} {\bibfnamefont
  {Z.}~\bibnamefont {Hatfield-Dodds}}, \bibinfo {author} {\bibfnamefont
  {R.}~\bibnamefont {Lasenby}}, \bibinfo {author} {\bibfnamefont
  {D.}~\bibnamefont {Drain}}, \bibinfo {author} {\bibfnamefont
  {C.}~\bibnamefont {Chen}}, \bibinfo {author} {\bibfnamefont {R.}~\bibnamefont
  {Grosse}}, \bibinfo {author} {\bibfnamefont {S.}~\bibnamefont {McCandlish}},
  \bibinfo {author} {\bibfnamefont {J.}~\bibnamefont {Kaplan}}, \bibinfo
  {author} {\bibfnamefont {D.}~\bibnamefont {Amodei}}, \bibinfo {author}
  {\bibfnamefont {M.}~\bibnamefont {Wattenberg}},\ and\ \bibinfo {author}
  {\bibfnamefont {C.}~\bibnamefont {Olah}},\ }\href@noop {} {\bibinfo {title}
  {{Toy Models of Superposition}}} (\bibinfo {year} {2022}),\ \Eprint
  {https://arxiv.org/abs/2209.10652} {arXiv:2209.10652 [cs.LG]} \BibitemShut
  {NoStop}%
\bibitem [{\citenamefont {Paulo}\ and\ \citenamefont
  {Belrose}(2025)}]{Paulo:2025same}%
  \BibitemOpen
  \bibfield  {author} {\bibinfo {author} {\bibfnamefont {G.}~\bibnamefont
  {Paulo}}\ and\ \bibinfo {author} {\bibfnamefont {N.}~\bibnamefont
  {Belrose}},\ }\href@noop {} {\bibinfo {title} {{Sparse Autoencoders Trained
  on the Same Data Learn Different Features}}} (\bibinfo {year} {2025}),\
  \Eprint {https://arxiv.org/abs/2501.16615} {arXiv:2501.16615 [cs.LG]}
  \BibitemShut {NoStop}%
\bibitem [{sup()}]{supp}%
  \BibitemOpen
  \href@noop {} {}\bibinfo {note} {Supplemental Material, included with this preprint as the
  ancillary file 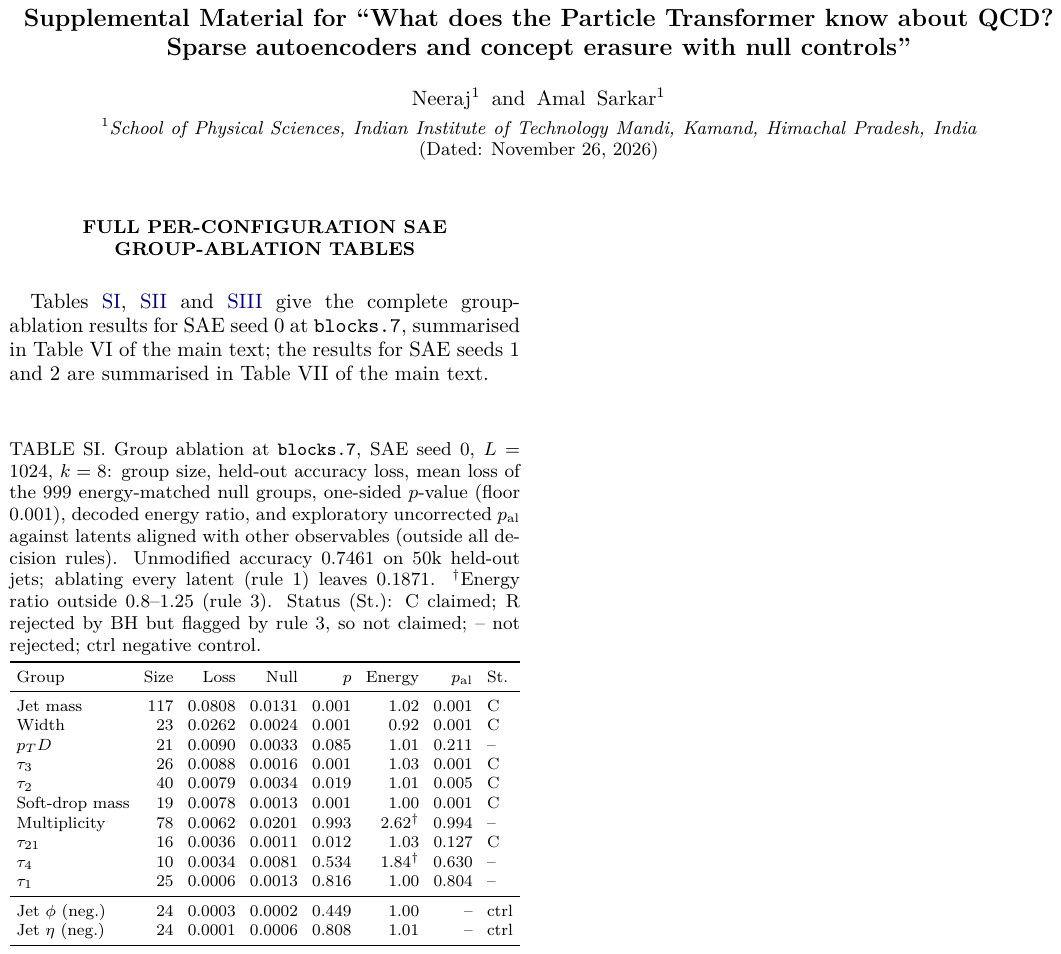: the complete per-configuration
  group-ablation tables}\BibitemShut {NoStop}%
\bibitem [{\citenamefont {Efron}(1979)}]{Efron:1979boot}%
  \BibitemOpen
  \bibfield  {author} {\bibinfo {author} {\bibfnamefont {B.}~\bibnamefont
  {Efron}},\ }\href {https://doi.org/10.1214/aos/1176344552} {\bibfield
  {journal} {\bibinfo  {journal} {Ann. Statist.}\ }\textbf {\bibinfo {volume}
  {7}},\ \bibinfo {pages} {1} (\bibinfo {year} {1979})}\BibitemShut {NoStop}%
\bibitem [{\citenamefont {Komiske}\ \emph {et~al.}(2019)\citenamefont
  {Komiske}, \citenamefont {Metodiev},\ and\ \citenamefont
  {Thaler}}]{Komiske:2018cqr}%
  \BibitemOpen
  \bibfield  {author} {\bibinfo {author} {\bibfnamefont {P.~T.}\ \bibnamefont
  {Komiske}}, \bibinfo {author} {\bibfnamefont {E.~M.}\ \bibnamefont
  {Metodiev}},\ and\ \bibinfo {author} {\bibfnamefont {J.}~\bibnamefont
  {Thaler}},\ }\href {https://doi.org/10.1007/JHEP01(2019)121} {\bibfield
  {journal} {\bibinfo  {journal} {J. High Energy Phys.}\ }\textbf {\bibinfo
  {volume} {01}}\bibfield  {number} {\bibinfo  {number} { (2019)},\ \bibinfo
  {pages} {121}},\ }\Eprint {https://arxiv.org/abs/1810.05165}
  {arXiv:1810.05165 [hep-ph]} \BibitemShut {NoStop}%
\bibitem [{\citenamefont {Konar}\ \emph {et~al.}(2022)\citenamefont {Konar},
  \citenamefont {Ngairangbam},\ and\ \citenamefont
  {Spannowsky}}]{Konar:2021zdg}%
  \BibitemOpen
  \bibfield  {author} {\bibinfo {author} {\bibfnamefont {P.}~\bibnamefont
  {Konar}}, \bibinfo {author} {\bibfnamefont {V.~S.}\ \bibnamefont
  {Ngairangbam}},\ and\ \bibinfo {author} {\bibfnamefont {M.}~\bibnamefont
  {Spannowsky}},\ }\href {https://doi.org/10.1007/JHEP02(2022)060} {\bibfield
  {journal} {\bibinfo  {journal} {J. High Energy Phys.}\ }\textbf {\bibinfo
  {volume} {02}}\bibfield  {number} {\bibinfo  {number} { (2022)},\ \bibinfo
  {pages} {060}},\ }\Eprint {https://arxiv.org/abs/2109.14636}
  {arXiv:2109.14636 [hep-ph]} \BibitemShut {NoStop}%
\bibitem [{\citenamefont {Gras}\ \emph {et~al.}(2017)\citenamefont {Gras},
  \citenamefont {H{\"o}che}, \citenamefont {Kar}, \citenamefont {Larkoski},
  \citenamefont {L{\"o}nnblad}, \citenamefont {Pl{\"a}tzer}, \citenamefont
  {Si{\'o}dmok}, \citenamefont {Skands}, \citenamefont {Soyez},\ and\
  \citenamefont {Thaler}}]{Gras:2017jty}%
  \BibitemOpen
  \bibfield  {author} {\bibinfo {author} {\bibfnamefont {P.}~\bibnamefont
  {Gras}}, \bibinfo {author} {\bibfnamefont {S.}~\bibnamefont {H{\"o}che}},
  \bibinfo {author} {\bibfnamefont {D.}~\bibnamefont {Kar}}, \bibinfo {author}
  {\bibfnamefont {A.}~\bibnamefont {Larkoski}}, \bibinfo {author}
  {\bibfnamefont {L.}~\bibnamefont {L{\"o}nnblad}}, \bibinfo {author}
  {\bibfnamefont {S.}~\bibnamefont {Pl{\"a}tzer}}, \bibinfo {author}
  {\bibfnamefont {A.}~\bibnamefont {Si{\'o}dmok}}, \bibinfo {author}
  {\bibfnamefont {P.}~\bibnamefont {Skands}}, \bibinfo {author} {\bibfnamefont
  {G.}~\bibnamefont {Soyez}},\ and\ \bibinfo {author} {\bibfnamefont
  {J.}~\bibnamefont {Thaler}},\ }\href
  {https://doi.org/10.1007/JHEP07(2017)091} {\bibfield  {journal} {\bibinfo
  {journal} {J. High Energy Phys.}\ }\textbf {\bibinfo {volume} {07}}\bibfield
  {number} {\bibinfo  {number} { (2017)},\ \bibinfo {pages} {091}},\ }\Eprint
  {https://arxiv.org/abs/1704.03878} {arXiv:1704.03878 [hep-ph]} \BibitemShut
  {NoStop}%
\bibitem [{\citenamefont {Bellm}\ \emph {et~al.}(2016)\citenamefont {Bellm}
  \emph {et~al.}}]{Bellm:2015jjp}%
  \BibitemOpen
  \bibfield  {author} {\bibinfo {author} {\bibfnamefont {J.}~\bibnamefont
  {Bellm}} \emph {et~al.},\ }\href
  {https://doi.org/10.1140/epjc/s10052-016-4018-8} {\bibfield  {journal}
  {\bibinfo  {journal} {Eur. Phys. J. C}\ }\textbf {\bibinfo {volume} {76}},\
  \bibinfo {pages} {196} (\bibinfo {year} {2016})},\ \Eprint
  {https://arxiv.org/abs/1512.01178} {arXiv:1512.01178 [hep-ph]} \BibitemShut
  {NoStop}%
\bibitem [{\citenamefont {Aad}\ \emph {et~al.}(2024)\citenamefont {Aad} \emph
  {et~al.}}]{ATLAS:2024rua}%
  \BibitemOpen
  \bibfield  {author} {\bibinfo {author} {\bibfnamefont {G.}~\bibnamefont
  {Aad}} \emph {et~al.} (\bibinfo {collaboration} {ATLAS}),\ }\href
  {https://doi.org/10.1088/1748-0221/19/08/P08018} {\bibfield  {journal}
  {\bibinfo  {journal} {JINST}\ }\textbf {\bibinfo {volume} {19}}\bibfield
  {number} {\bibinfo  {number} { (2024)},\ \bibinfo {pages} {P08018}},\
  }\Eprint {https://arxiv.org/abs/2407.20127} {arXiv:2407.20127 [hep-ex]}
  \BibitemShut {NoStop}%
\bibitem [{\citenamefont {Tumasyan}\ \emph {et~al.}(2022)\citenamefont
  {Tumasyan} \emph {et~al.}}]{CMS:2021iwu}%
  \BibitemOpen
  \bibfield  {author} {\bibinfo {author} {\bibfnamefont {A.}~\bibnamefont
  {Tumasyan}} \emph {et~al.} (\bibinfo {collaboration} {CMS}),\ }\href
  {https://doi.org/10.1007/JHEP01(2022)188} {\bibfield  {journal} {\bibinfo
  {journal} {J. High Energy Phys.}\ }\textbf {\bibinfo {volume} {01}}\bibfield
  {number} {\bibinfo  {number} { (2022)},\ \bibinfo {pages} {188}},\ }\Eprint
  {https://arxiv.org/abs/2109.03340} {arXiv:2109.03340 [hep-ex]} \BibitemShut
  {NoStop}%
\bibitem [{\citenamefont {Aad}\ \emph {et~al.}(2016)\citenamefont {Aad} \emph
  {et~al.}}]{ATLAS:2016vxz}%
  \BibitemOpen
  \bibfield  {author} {\bibinfo {author} {\bibfnamefont {G.}~\bibnamefont
  {Aad}} \emph {et~al.} (\bibinfo {collaboration} {ATLAS}),\ }\href
  {https://doi.org/10.1140/epjc/s10052-016-4126-5} {\bibfield  {journal}
  {\bibinfo  {journal} {Eur. Phys. J. C}\ }\textbf {\bibinfo {volume} {76}},\
  \bibinfo {pages} {322} (\bibinfo {year} {2016})},\ \Eprint
  {https://arxiv.org/abs/1602.00988} {arXiv:1602.00988 [hep-ex]} \BibitemShut
  {NoStop}%
\bibitem [{\citenamefont {Aad}\ \emph {et~al.}(2020{\natexlab{a}})\citenamefont
  {Aad} \emph {et~al.}}]{ATLAS:2020bbn}%
  \BibitemOpen
  \bibfield  {author} {\bibinfo {author} {\bibfnamefont {G.}~\bibnamefont
  {Aad}} \emph {et~al.} (\bibinfo {collaboration} {ATLAS}),\ }\href
  {https://doi.org/10.1103/PhysRevLett.124.222002} {\bibfield  {journal}
  {\bibinfo  {journal} {Phys. Rev. Lett.}\ }\textbf {\bibinfo {volume} {124}},\
  \bibinfo {pages} {222002} (\bibinfo {year} {2020}{\natexlab{a}})},\ \Eprint
  {https://arxiv.org/abs/2004.03540} {arXiv:2004.03540 [hep-ex]} \BibitemShut
  {NoStop}%
\bibitem [{\citenamefont {Aad}\ \emph {et~al.}(2020{\natexlab{b}})\citenamefont
  {Aad} \emph {et~al.}}]{ATLAS:2019mgf}%
  \BibitemOpen
  \bibfield  {author} {\bibinfo {author} {\bibfnamefont {G.}~\bibnamefont
  {Aad}} \emph {et~al.} (\bibinfo {collaboration} {ATLAS}),\ }\href
  {https://doi.org/10.1103/PhysRevD.101.052007} {\bibfield  {journal} {\bibinfo
   {journal} {Phys. Rev. D}\ }\textbf {\bibinfo {volume} {101}},\ \bibinfo
  {pages} {052007} (\bibinfo {year} {2020}{\natexlab{b}})},\ \Eprint
  {https://arxiv.org/abs/1912.09837} {arXiv:1912.09837 [hep-ex]} \BibitemShut
  {NoStop}%
\bibitem [{\citenamefont {Phipson}\ and\ \citenamefont
  {Smyth}(2010)}]{Phipson:2010perm}%
  \BibitemOpen
  \bibfield  {author} {\bibinfo {author} {\bibfnamefont {B.}~\bibnamefont
  {Phipson}}\ and\ \bibinfo {author} {\bibfnamefont {G.~K.}\ \bibnamefont
  {Smyth}},\ }\href {https://doi.org/10.2202/1544-6115.1585} {\bibfield
  {journal} {\bibinfo  {journal} {Stat. Appl. Genet. Mol. Biol.}\ }\textbf
  {\bibinfo {volume} {9}},\ \bibinfo {pages} {39} (\bibinfo {year}
  {2010})}\BibitemShut {NoStop}%
\end{thebibliography}%

\end{document}